**A Tale of Two Cities: The Announcement Effect of Northern Metropolis Plan**

Yi Fan[a], Chongyu Wang[b], Ke Xu[c]

[a] Corresponding author. Department of Real Estate, NUS Business School, National University of Singapore. 15 Kent Ridge Dr., Singapore 119245. yi.fan@nus.edu.sg. Yi Fan acknowledges the Ministry of Education Singapore Academic Research Fund A-8002711-00-00.

[b] Department of Risk Management/Insurance, Real Estate & Legal Studies, Florida State University. 402 W. Gaines Street, Tallahassee, FL 32301. cwang@wertheim.fsu.edu

[c] Department of Real Estate and Construction, Knowles Building, University of Hong Kong. Pokfulam Road, Hong Kong. xkelly103@connect.hku.hk

**Abstract**

Using the October 2021 announcement of Hong Kong's Northern Metropolis Plan as a quasi-natural experiment, we examine its impacts on households, firms, and demographics. Leveraging 100,576 housing transactions, 4.7 million consumption records, and district-level demographic data, difference-in-differences estimates show housing prices in the treated region rose by 3.9% within one year, alongside a 4.1% increase in consumption. Firm entry expanded, with inflows of higher-income, higher-educated households. We find a mild, lagged spillover to Shenzhen's housing market but no significant firm response. While the policy narrows cross-region inequality, it increases within-region inequality and raises affordability concerns for lower-SES households.

**Keywords:** Place-based Policy, Housing Prices, Online Consumption, Firm Establishment, Demographic Changes, Regional Spillover Effect
**JEL Classification:** R23, R38, R58, L81

## 1. Introduction

Place-based policies are government efforts designed to foster economic development and well-being within specific geographic areas (Freedman and Neumark, 2024). These policies, often involving large-scale public spending and government efforts, have become pivotal to stimulating private sector investment and economic growth (Duranton and Venables, 2018). While existing research typically evaluates their aggregate impacts on jobs and the economy, it often overlooks the micro-level behavioral responses of market participants. Specifically, little is known about how people respond to policy announcements at the granular level—adjusting their household consumption, business establishment activities or migration decisions even before any physical infrastructure materializes (Agrawal et al., 2022; LaPoint and Sakabe, 2021). However, it is crucial to understand how such policy blueprints shift demand for local amenities and housing, which are important for later policy adjustments. Moreover, the effects of place-based policies can spill over beyond target zones, influencing labor mobility, housing markets, and industrial linkages in a broader urban network (Neumark and Simpson, 2015; Lu et al., 2019).

We use the announcement of the Northern Metropolis Plan in Hong Kong as a quasi-natural experiment to bridge this understanding gap. On 6 October 2021, Carrie Lam, the Chief Executive of Hong Kong, unveiled an ambitious plan to transform two remote districts, North and Yuen Long—accounting for 27% of Hong Kong's total area and bordering Shenzhen, China's third-largest city after Shanghai and Beijing—into Northern Metropolis. The long-term aspirations are to alleviate the city's notorious housing affordability crisis and foster a new high-tech hub. Notably, Hong Kong has championed the world's least affordable housing market for an unprecedented 14$^{th}$ year. Providing affordable public housing to vulnerable residents is a remedy to cure the unrest. It is projected to build an additional 500,000 housing units to accommodate 1.5 million more residents (i.e., 1/5 of the total population) and create 500,000 new jobs, including 150,000 positions in the IT sector (Marlow et al., 2021) (policy details in Section 2). The enactment of the Northern Metropolis policy provides a valuable policy arena to estimate how a massive urban development plan can alter human behaviors and affect regional economic development, both in the short term and over the long haul.

Equipped with highly dynamic, granular datasets, this paper examines how households and firms respond to the announcement of the Northern Metropolis plan—through changes in housing demand, consumption, migration, and firm activities. We also consider its potential spillover effects to neighboring Shenzhen, shedding light on how place-based policies can reshape regional dynamics beyond their intended boundaries. In addition to short-term analyses, we also provide longer-term predictions, using the Cobb-Douglas production function and the gravity model of migration to conduct a back-of-the-envelope calculation of the Plan's longer-term socioeconomic implications (Cobb and Douglas, 1928).

We begin by assessing the impact of the policy on household behavior, using a difference-in-differences (DiD) framework that leverages the announcement of the policy as an exogenous shock. The North and Yuen Long districts serve as the treatment group, while the remaining 16 districts constitute the control group. To measure shifts in households' home purchasing decisions, we analyze 100,576 housing transaction records spanning October 2020 to October 2022. Our results show a significant 3.9% increase in average transaction prices within the treatment group one year after the policy announcement. The most pronounced rise of 6.9% occurred within four months after the announcement. For context, the average annualized return on private residential property in Hong Kong from 2015 to 2025 is merely 1.96%.

Next, we examine household consumption behavior following the announcement. Literature shows that household consumption is often excessively sensitive to macroeconomic news (Garmaise et al., 2024). Drawing from a rich dataset of approximately 4.7 million transaction records from Hong Kong's leading online shopping platform, we observe that the transaction amount per order placed by customers within the Northern Metropolis areas surged by nearly 4.1% after the policy was enacted. This rise in consumption closely mirrors the estimated increase in housing values, consistent with the Permanent Income Hypothesis (PIH). Intriguingly, this consumption buoyancy is concentrated among residents of private housing estates, who are likely more affluent and responded more strongly to perceived gains in housing wealth by refining their consumption preferences in the policy's wake.

On the firm side, we observe a rapid increase in firm establishments. Within four years of the announcement, the number of registered establishments rose significantly by 14.4% in the information and communications (I&C) sector, and by 25.7% in the construction sector. Because only two of eighteen districts are treated, we supplement our district-level analyses with the Synthetic Difference-in-Differences (SDiD) estimation (Arkhangelsky et al., 2021), which reweights control units to construct a more credible counterfactual. The SDiD estimates are closely aligned with our baseline results across all outcomes. Together, these patterns highlight the sensitive responses of business activities to policy-oriented local economic opportunities.

We further delve into the policy impact on demographic composition. Following the announcement, the share of higher-educated residents within the Northern Metropolis surged by 6.54%. The number of higher-income and family-oriented households increased by 9.04% and 5.22%, respectively. These shifts suggest that place-based policies can prompt forward-looking household decisions and offer a lens into how place-based policies may shape intergenerational outcomes (Black et al., 2020; Agarwal et al., 2026; Daysal et al., 2023; Yu et al., 2023; Weiwu, 2025).

No local policy exists in a vacuum; instead, its outcomes are shaped not only by direct effects but also by interactions with neighboring jurisdictions (Agrawal et al., 2022). Located within the Guangdong-Hong Kong-Macao Greater Bay Area (GBA), China's dynamic southern megaregion and the world's 10$^{th}$ largest economy, the announcement of the Northern Metropolis likely has cross-border spillovers. We find evidence of modest increases in housing prices and transaction volume in Shenzhen following the announcement, with a lag of approximately three months. This spillover effect suggests the policy is reshaping economic geography by extending its impact beyond Hong Kong's administrative boundaries.

Our study is inconclusive on inequality. Hong Kong is formidable in its rising inequality. With 125,100 millionaires, the city has 1.6 million people living in poverty (Yeung, 2022). The recent unveiling of the Northern Metropolis Plan has yielded significant uplifts in housing prices, especially for homeowners, household consumption, and demographic changes, foreshadowing enduring regional prosperity through rising inequality. Yet, as the disparity between the Northern Metropolis and city centers wanes, the inequality within the Northern Metropolis might amplify. Spiraling housing prices could exacerbate wealth gaps, particularly dichotomies between private and public housing occupants. The unaffordability within the region may deteriorate, pushing the vulnerable at the city border, geographically or economically, to become even more vulnerable. Nevertheless, fueled by the large-scale public housing construction and an influx of higher-educated and higher-income skilled labor in the high-tech sector, the housing unaffordability is likely to be mitigated, and a potent multiplier effect could emerge, promising ripple benefits in the non-traded sectors, welfare benefits to the locals and migrants, intra- and inter-generationally.[1]

Our paper contributes to the literature from at least two aspects. First, we investigate the announcement effect of a very recent government-led mega project, which is poised to reshape the lives of more than 1 million present inhabitants and to draw 1/3 of the population from congested city centers in two decades. To the best of our knowledge, we are the first to provide a holistic view of the effect of the Northern Metropolis plan on household and firm behaviors and regional spillovers. Different from literature focusing on post-implementation impacts of mega projects or place-based policies (Moran et al., 2018; Basheer et al., 2021; Vilela et al., 2020; Diao et al., 2021; Lin et al., 2021; Holtgrieve and Arias, 2022; Galelli et al., 2022), our analysis of the announcement effect offers a forward-looking lens for policy implementation and adjustment, fostering a future-proof city.

Policy wise, the Northern Metropolis plan is parallel to the Economic Zone Program in P.R. China (Lu et al., 2019), the Industrial Transfer Policy (Gerritse et al., 2026), the Opportunity Zones in the U.S. (Chen, Glaeser, et al., 2023), the Enterprise Zone policies in France (Neumark and Simpson,

[1] Notably, the multiplier effect of high-tech employment is three times larger than that in traditional manufacturing industries (Moretti and Thulin, 2013). Additionally, skilled workers are more innovative and react more quickly to economic shocks, and thus are expected to contribute to the reinvention of cities (Glaeser et al., 2004).

2015), Urban Revitalization Program in the Netherlands (Koster and Ommeren, 2019), Initiative for the Integration of Regional Infrastructure in South America (Vilela et al., 2020), and megaprojects such as building large dams and oil sands mining (Rooney et al., 2015; Moran et al., 2018). In recent years, the adoption of industrial policies—a set of specific policy tools designed to bolster sectors deemed strategically important—has become common as means to tackle market failures and promote rapid, broadly shared economic growth. This raises the pertinent question: What is the general equilibrium impact of a place-based policy located in a vibrant economic zone? Our findings suggest that the Northern Metropolis policy, particularly the contents favoring the high-tech industry like the San Tin Technopole, has the potential to attract skilled labor and impact regional economic development, but may come at the expense of exacerbating inequalities. Future studies are warranted along this line with the progress of the Northern Metropolis redevelopment.

## 2. Institutional Background

Hong Kong is now regarded as the most expensive city to live in globally (Cheshire et al., 2018). According to the 2021 edition of the Demographia International Housing Affordability Survey, Hong Kong ranked as the 1st least affordable metropolitan market worldwide, with a median multiple of 20.7, almost four times the threshold for the definition of "severely unaffordable".[2][3] Hong Kong is a highly developed city. However, for historical reasons, there is still a lot of undeveloped land in the northern part of Hong Kong, near Shenzhen, China.

To spur the development of a new economy in the northern part of Hong Kong, solve the housing affordability issue in Hong Kong, and strengthen the integration of the Greater Bay Area, on October 6th, 2021, the Hong Kong government unexpectedly announced the "Northern Metropolis Development Strategy". The Northern Metropolis covers the northern part of Hong Kong, not only including some established new towns in that area, such as Fanling, Tin Shui Wai, and Sheung Shui, etc., but also including some undeveloped areas, such as Hung Shui Kiu, Kwu Tung North, and San Tin, etc., with a total area of 30,000 hectares.

[2] The median multiple is the ratio of the median home price to the median annual household income.

[3] The international standard for unaffordable housing price is reached when the median multiple exceeds 5.5.

The strategy includes four pillars: 1) The government plans major transportation development, including five new railways, three of which are cross-border, to enhance linkages between Hong Kong and Shenzhen. 2) The plan will expand land supply for innovation and technology uses, centered on the San Tin Technopole, a roughly 1,100-hectare hub designed to support I&T enterprises and research institutes. 3) The plan will substantially increase the housing supply through several large development projects across the Northern Metropolis, with new residential projects in both the western part of the Northern Metropolis (Lau Fau Shan and Tsim Bei Tsui) and the eastern part of the Northern Metropolis (Lo Wu and Man Kam To). 4) Finally, the government aims to enhance environmental capacity by resuming private wetlands and fishponds, expanding wetland conservation parks, and developing urban-rural greenways. Details of the policy are presented in Appendix 1.

## 3. Data and Sample

Our analysis combines multiple datasets covering housing markets, consumption behavior, firm activity, and demographics in Hong Kong and Shenzhen. Housing transaction data for Hong Kong is obtained from EPRC Ltd., a widely used and esteemed reservoir of housing transaction data. The dataset contains transaction-level details such as transaction price, date, buyer and seller names, property address, building age, floor level, saleable floor area, and the number of rooms. We define the Northern Metropolis area as comprising the North and Yuen Long districts, two of Hong Kong's eighteen districts.

To measure consumption behavior, we use deal-level online consumption records from HKTV Mall, the largest online shopping platform in Hong Kong. This dataset spans June 2021 to February 2022 and contains 4,686,504 observations.[4] Unlike traditional Census data, which are low frequency

[4] According to an online retail survey from Frost & Sullivan in 2021, the HKTV Mall leads the online retail landscape in Hong Kong with a commanding 13.9 percent market share. Its closest competitor, PARKnSHOP, lags significantly, holding a market share of merely 2.4 percent of the market, showcasing HKTV Mall's dominance. See: Yoho Group Holdings Limited Prospectus, available at https://www1.hkexnews.hk/listedco/listconews/sehk/2022/0610/10298050/sehk22052401442.pdf

(e.g., every 10 years) and not viable for evaluating short-term policy effects, HKTV Mall provides a high-frequency, granular perspective of consumption behavior. Each transaction includes buyer ID, order amount, order date, delivery area, building name, and housing type. Importantly, housing type allows us to distinguish between private housing and public housing (e.g., the Home Ownership Scheme and public rental housing); for the latter resale prices are often unavailable.

Firm establishment data are sourced from the Hong Kong Census and Statistics Department and cover quarterly observations from 2015 to 2025. This dataset includes the number of registered establishments and employees by industry at the district level, allowing us to track changes in business activity and industry composition over time. We focus on the information and communication sector and the construction sector, which are directly targeted by the Northern Metropolis plan. In addition, we collect annual district-level demographic data for 2017-2023 from the Hong Kong Census and Statistics Department. Key variables include the percentage of residents with university education, the number of households with a monthly income above HKD 30,000 and the number of households with two or more members.

To examine cross-border effects, we use a proprietary housing transaction data for Shenzhen with variables comparable to the EPRC data. These include price, date, address, building age, floor level, unit size, and elevator availability. We define Shenzhen's bordering area as the five subdistricts affected by the Northern Metropolis area: Yuehai, Nantou, Futian, Fubao, and Huangbei.[5] Firm establishment data for Shenzhen is drawn from the State Administration for Industry and Commerce (SAIC), following Chen, Cheng, et al. (2023).

---

[5] For the transportation development in Northern Metropolis Plan, firstly, Hong Kong government will construct Hong Kong-Shenzhen western rail to upgrade the transport linkage with the Qianhai area in Shenzhen. This project will impact two metro stations in Shenzhen: Shenzhen Bay Port station (located in *Yuehai* subdistrict) and Qianhai Bay station (located in *Nantou* subdistrict). Secondly, Hong Kong government will construct the northern link spur line to link the Hong Kong-Shenzhen Innovation and Technology Park (adjacent to *Futian* subdistrict) and new Huanggang port (located in *Fubao* subdistrict). Thirdly, Hong Kong government will construct the northern link eastward extension to develop Man Kam To area (adjacent to *Huangbei* subdistrict). China's urban governance follows a top-down hierarchy: municipality, district (区, qū), subdistrict (街道, jiēdào), and community. A district is a major administrative unit, akin to a borough or large U.S. county, while a subdistrict is a lower-level division responsible for neighborhood-level administration through a Subdistrict Office (街道办事处, jiēdào bànshìchù). For example, the treated subdistricts *Yuehai* and *Nantou* fall under Nanshan district, *Futian* and *Fubao* belong to Futian district and *Huangbei* belongs to Luohu district.

Descriptive statistics for all samples are presented in Appendix Table A1 (variable definitions are provided in Appendix 2). A median housing transaction in Hong Kong takes place on the 15$^{th}$ floor of an eighteen-year-old property. An average unit transacted has a market value of HKD 7.594 million (approximately USD 0.97 million) and a size of 492 square feet (Panel A).[6] The median online shopping per order is HKD 474.5 (USD 60.45). The mean of *North* is 0.105, indicating that customers living outside the Northern Metropolis use online consumption more frequently (Panel B). On the firm side, the median numbers of establishments for the I&C and construction industry are 681 and 82 per district, respectively (Panel C). Regarding Hong Kong's demographics, over a quarter of the population (27.2%) holds a bachelor's degree or above, reflecting a relatively high level of educational attainment. The median numbers per district of higher-income and family-oriented households (with more than 2 household members) are 64,000 and 118,050, respectively (Panel D).

Across the border, a median housing transaction in Shenzhen takes place on the 11$^{th}$ floor of an 18-year-old building, which is similar to Hong Kong. However, the median home price is CNY 4.088 million (USD 0.59 million) with a unit size of 778.076 square feet, indicating more affordable and spacious housing. About a quarter of transactions (0.261) are located near the Hong Kong–Shenzhen border, suggesting strong housing demand in Shenzhen for properties close to Hong Kong (Panel E). On the firm side of Shenzhen, the median number of establishments in the I&C industry is 39 per subdistrict, whereas merely 5 in the construction industry.

**4. Empirical Models**

We estimate the announcement effects of the Northern Metropolis policy using a DiD framework:

$$Y_{i(l)t} \ = \ \alpha + \beta_1 Policy_t * Treat_i + \gamma X_{it} + \varphi_{lt} + \omega_i + \varepsilon_{i(l)t} \qquad (1)$$

where $Y_{i(l)t}$ denotes the outcome of interest for unit *i* at time *t*. Outcomes include housing prices, online consumption, firm establishments, employment and demographic composition. $Policy_t$ is an indicator that equals 1 for periods after October 2021 and 0 otherwise. $Treat_i$ indicates exposure to the policy and varies by analysis, such as properties or residents in the North and Yuen Long districts in the specifications for Hong Kong's housing transactions and consumption behaviors, firms located in the Northern Metropolis in the specification for Hong Kong's firm establishments, or subdistricts adjacent to border in the specification estimating spillover effects in Shenzhen. The coefficient $\beta_1$ captures the average treatment effect of the Northern Metropolis announcement on the respective outcome of interest.

[6] The USD/HKD exchange rate on the policy announcement date was 7.85. We use this rate consistently for all currency conversions during this study (https://www.hkma.gov.hk/eng/data-publications-and-research/data-and-statistics/daily-monetary-statistics/2021/10/ms-20211006/).

$X_{it}$ encompasses a vector of time-varying controls when applicable. For housing outcomes, these controls include building age, floor level, and unit size. For demographic changes, we control for the district-level median monthly household income. $\varphi_{lt}$ represents time fixed effects or time-by-location fixed effects, depending on the empirical setting. $\omega_i$ denotes unit fixed effects, such as customer, district, or subdistrict fixed effects. The error term is $\varepsilon_{i(l)t}$. Standard errors are clustered at the time and location level appropriate for the sample employed.

## 5. Main Results

### 5.1 Housing

As discussed, place-based policies can influence economic outcomes by shaping expectations and triggering behavioral responses among households (Duranton and Venables, 2018; Bilal and Rossi-Hansberg, 2021). These effects tend to be more pronounced in areas with inelastic housing supply, where projected future gains are quickly capitalized into land values or housing costs (Austin et al., 2018).

The plausibly exogenous nature of the Northern Metropolis policy shock is supported by the unexpected timing of the announcement, which occurred during the leadership transition between chief executives. This reduces the likelihood that the announcement was anticipated or strategically timed relative to local market conditions. To further validate our identification strategy, we test and confirm the parallel trends assumption: before the policy, housing prices in the Northern Metropolis area and the rest of Hong Kong had trended similarly. In contrast, we observed a clear divergence in post-announcement housing price trajectories (Figure 1).

[INSERT FIGURE 1 ABOUT HERE]

We then quantify how housing prices responded to the policy announcement. On average, housing prices in the Northern Metropolis area increased by 7.6% (HKD 733,780 or USD 93,475) within four months and by 5% (482,750 HKD or 61,497 USD) within one year of the announcement, controlling for housing features, year-month, and district fixed effects (Column (1)-(2), Table 1). To further account for unobserved factors that may vary across time and regions, we include year-month-by-region fixed effects in the last two columns, which we treat as our preferred specification. The corresponding point estimate remains significant at the 1% level, though slightly attenuated to 6.9% and 3.9% (HKD 666,195 or USD 84,866 and HKD 376,545 or USD 47,968 for an average home, respectively (Column (3)-(4), Table 1).

The magnitude and timing of these price changes suggest households rapidly revised their expectations in response to the announcement of the policy. These early adjustments in willingness to pay—well in advance of any physical development—reflect the forward-looking nature of market participants and future payoffs associated with improved housing amenities, connectivity, and neighborhood value. Our results remain robust when restricting the analysis to transactions within the New Territories and winsorizing the top and bottom 1% of observations (see Appendix Table A2). Alternatively, we use monthly building-level aggregated transaction amounts as the outcome variable (Appendix Figure A1 and Appendix Table A3). These results reinforce our findings in Table 1, indicating that the announcement of the Northern Metropolis policy not only drove up housing prices but also stimulated transaction activity in the area.

[INSERT TABLE 1 ABOUT HERE]

**5.2 Consumption**

With evident rise in housing prices, we examine how the policy announcement affected consumption possibly through housing wealth effects or tighter household budget constraints arising from higher housing costs (Ding et al., 2025). To distinguish these competing effects, we separate households in private housing (who benefit from housing wealth gains) and those in public housing (who face cost pressure without the gains due to resale restrictions).

Using the DiD framework outlined in Equation (1), we find a significant surge in online shopping in the Northern Metropolis area. As shown in Figure 2, spending in the treatment group rose sharply relative to other districts approximately four months after the policy announcement, despite parallel trends prior to the announcement.

[INSERT FIGURE 2 ABOUT HERE]

Table 2 presents DiD estimates based on our full sample, as well as subsets of order volume for private and public housing, respectively. In addition to the year-month and district fixed effects specified in Table 1, we further include customer fixed effects to account for unobserved, time-invariant heterogeneity across consumers. We find that the shopping expenditure per order within the Northern Metropolis area rose by 4.1% relative to the rest of Hong Kong following the policy announcement, as revealed by our preferred estimates presented in Column (4). This effect is primarily driven by private housing occupants (Column (5)), whose order volume rose by 4.7%, while the increase for public housing residents is modest (Column (6)). The difference between the two point estimates is statistically significant at the 1% level. These results remain robust when restricting the sample to orders within the New Territories and winsorizing the top and bottom 1% of observations (see Appendix Table A4). The stronger response among private housing residents suggests that anticipated gains in housing wealth led to increased consumption, consistent with the permanent income hypothesis.

[INSERT TABLE 2 ABOUT HERE]

We further conduct heterogeneity analysis by decomposing consumption into three categories: necessities, non-necessities, and entertainment (Appendix Table A5). Private estate occupants in the Northern Metropolis area increase their spending across all three categories after the policy announcement, whereas public housing occupants only spend moderately more on necessities. Again, it is consistent with the main finding that while private estate occupants may capitalize on rising property values by selling their homes and realizing capital gains, public housing occupants face stringent resale restrictions that likely limit their ability to realize capital gains and smooth consumption.[7]

### 5.3 Productivity

Our analysis so far focuses on household responses to the Northern Metropolis plan. We now turn to firm behavior to examine whether the policy announcement operates through a "Big Push" on the production side (Glaeser and Gottlieb, 2008; Kline and Moretti, 2014). Specifically, in the presence of agglomeration forces, a large, coordinated policy intervention can overcome coordination problems and shift a region from a low-productivity equilibrium to a high-productivity one by simultaneously attracting firms and workers. Importantly, recent evidence suggests these agglomeration forces disproportionately benefit high-technology industries through mechanisms such as knowledge spillovers and talent clustering (Jiang et al., 2025). We therefore examine changes in business activity in I&C technology and construction sectors as key indicators of policy effectiveness. This also helps dissect the household responses by clarifying the types of workers sorting into the target region.

Figure 3 depicts the time-varying policy impact on firm establishments in the I&C industry (Panel A) and construction industry (Panel B), respectively. In both cases, coefficients prior to the policy announcement are statistically indistinguishable from zero, supporting that the Northern Metropolis area and other regions follow parallel trends before the policy announcement. Following the announcement, there is a significant surge in firm establishments in the Northern Metropolis area relative to other districts.

[7] We classify online consumption into two main categories: necessities (i.e., housewares, gadgets, and supermarket goods) and non-necessities (i.e., clothes, makeup, skincare, travel, sports, toys, and pets). We further divide non-necessities into entertainment-related (i.e., travel, sports, toys, and pets) versus other discretionary spending. Our results remain consistent when using a refined counterfactual and winsorizing key variables (Appendix Table A6).

[INSERT FIGURE 3 ABOUT HERE]

Table 3 presents DiD estimates of the policy impact on firm establishments and employment at the district-quarter level. Columns (1) and (2) show that the number of firm establishments and employees in the I&C industry increased by 14.42% and 11.73%, respectively. The construction industry exhibits even stronger responses, with establishments and employees rising by 25.72% and 35.57%. These patterns are consistent with the "Big Push" theory in which policy-induced improvements in infrastructure and housing supply attract complementary firm entry and labor demand in targeted sectors.

A potential concern with our DiD design is that only two of Hong Kong's eighteen districts, North and Yuen Long, constitute the treated group. This asymmetry between treated and control groups can affect estimation in two ways. First, when the control group is large and heterogeneous, the simple average of all control districts may provide a poor counterfactual for the treated districts, biasing the estimated treatment effect if some control districts follow systematically different trends. Second, with only two treated districts, conventional cluster-robust standard errors can suffer from finite-sample bias, potentially leading to over-rejection of the null hypothesis (Cameron and Miller, 2015). To address these concerns, we implement the Synthetic Difference-in-Differences (SDiD) estimator proposed by Arkhangelsky et al. (2021). The SDiD estimator combines the strengths of the difference-in-differences and synthetic control approaches: it reweights control units to better match the pre-treatment trajectory of the treated districts, constructing a more credible counterfactual while retaining the double-differencing structure that accounts for time-invariant unobservables.

The SDiD results, reported in Appendix Table A7, are consistent with our baseline findings. Firm establishments in the I&C and construction industries increase by 14.8% and 30.4%, respectively, and both estimates remain statistically significant. Comparing these magnitudes with our baseline DiD estimates of 14.4% and 25.7% (Table 3), the SDiD point estimates are closely aligned for the I&C sector and moderately larger for construction. The similarity between the two sets of estimates suggests that the composition of the control group does not materially distort our baseline results, and that the estimated treatment effects are not driven by a poor counterfactual arising from the unweighted average of all control districts.

[INSERT Table 3 ABOUT HERE]

### 5.4 Demographic Changes

Alongside changes in housing prices, consumption patterns, and firm establishments, the announcement of the Northern Metropolis Plan is also expected to reshape local demographics. Literature shows that place-based policies can influence human capital accumulation, both through direct investments in education and indirectly by improving neighborhood environments in ways that shape migration decisions (Agrawal et al., 2022; Weiwu, 2025). Improved access to schools, rising income and employment prospects, and expectations of local growth can encourage family relocation and greater educational attainment, particularly among upwardly mobile households. However, these benefits are often uneven, as more advantaged households are better positioned to capitalize on such opportunities. We examine whether the Northern Metropolis policy has induced early shifts in household composition, especially among more educated and higher-income households and families with school-aged children, that may signal emerging trends in human capital accumulation.

We observe a post-announcement inflow of higher-educated individuals, higher-income households, and households with children into the Northern Metropolis. These patterns are visually evident in Figure 4, which validates parallel pre-trends between the Northern Metropolis and other districts, and shows sharp post-announcement divergence.[8]

[INSERT FIGURE 4 ABOUT HERE]

Table 4 presents DiD estimates. Consistent with a forward-looking response, the Northern Metropolis area witnessed a 6.5% increase in the share of university educated residents, a 9.0% rise in higher-income households, and a 5.2% increase in households with larger family sizes. These shifts indicate that the policy may have attracted relatively affluent and family-oriented households into the region.

[8] The policy announcement overlaps with a period of elevated overseas emigration from Hong Kong; our difference-in-differences design mitigates this concern by comparing Northern Metropolis districts to other Hong Kong districts subject to similar aggregate demographic trends.

We apply the same SDiD approach to the demographic outcomes to address the small-cluster concern discussed above. As reported in Appendix Table A8, the SDiD estimates remain statistically significant across all three outcomes: the share of university-educated residents increases by 7.2%, higher-income households by 8.7%, and households with larger family sizes by 4.0%. These magnitudes are broadly consistent with the corresponding DiD estimates of 6.5%, 9.0%, and 5.2% (Table 4), with the SDiD point estimates slightly smaller for higher-income households and family size but marginally larger for education.

Taken together, these findings suggest that the Northern Metropolis is becoming increasingly attractive to a more educated, affluent, and family-oriented population following the announcement. This early demographic shift signals a growing perception of the area as a vibrant place for long-term residence and child-rearing. If sustained by coordinated investments in housing supply, educational capacity, and transportation, these changes could lay the groundwork for economic prosperity and social mobility.

[INSERT TABLE 4 ABOUT HERE]

### 5.5 Regional Spillover

Cities and regions are economically integrated systems. The effects of place-based policies often extend beyond their target boundaries, via spatial displacement, agglomeration spillovers, labor market integration, and capitalization effects (Neumark and Simpson, 2015; Corinth and Feldman, 2024; Fajgelbaum and Gaubert, 2025) (detailed channel discussion in Appendix 3). Theory offers two competing predictions: if policies merely displace activity, gains in treated areas should be offset by losses nearby. In contrast, agglomeration spillovers imply positive-sum effects, with rising demand and prices both inside and outside the treated area, potentially with a lag. The Northern Metropolis, located at the northern edge of Hong Kong and adjacent to downtown Shenzhen, provides a natural setting to examine such cross-border effects.

We examine cross-border spillovers by comparing housing market responses between five Shenzhen border subdistricts directly impacted by the Northern Metropolis area (Yuehai, Nantou, Futian, Fubao, and Huangbei) and other subdistricts within Nanshan, Futian, and Luohu districts, as discussed in Section 3. Figure 5 supports the parallel trends assumption. Unlike the immediate housing price response within the Northern Metropolis area, housing prices in Shenzhen adjust more gradually, with increases beginning approximately three months after the announcement. Using a DiD framework, we find that housing prices in Shenzhen's border subdistricts rose by 3.68% within twelve months of the announcement (Table 5, Column (4)). We find similar evidence when using transaction volume aggregated at the building-month level as the outcome variable (Appendix Figure A2 and Appendix Table A9).

[INSERT FIGURE 5 ABOUT HERE]

[INSERT TABLE 5 ABOUT HERE]

In contrast, we find no evidence of an increase in firm establishments in Shenzhen following the announcement (Appendix Table A10). Taken together, these results indicate that cross-border spillovers primarily operate through housing demand and expectation-driven price adjustments, rather than through a zero-sum displacement of economic activity. The delayed but significant response in Shenzhen highlights the importance of interregional spillovers when evaluating large place-based policies.

**6. Longer-Term Prediction**

To complement the short-term evidence, we also provide a set of long-run, model-based predictions for the Northern Metropolis plan. Specifically, we assume that the policy shock acts as a permanent increase in capital input (via infrastructure investment) and labor input (via the 500,000 new jobs), which we feed into a calibrated Cobb-Douglas production function combined with a gravity model of migration. We focus on four key dimensions: economic development, income growth, housing affordability, and migration. Our estimations assume that all planned developments in the Northern Metropolis area will be fully realized. Appendix Figure A3 presents a back-of-the-envelope calculation on the longer-term impact of the Northern Metropolis Plan upon completions: it is expected to generate GDP of USD 162.9 billion, increase the monthly wage by USD 712 dollars in the area, boost local housing prices by a moderate 5.9%, and attract at least 81,251 migrants. Appendix 4 details the steps for generating each prediction, and we caution that those estimates are rough and subject to changing socioeconomic conditions.

**7. Discussion and Conclusion**

This study offers a meticulous assessment of the Northern Metropolis policy, a major place-based intervention announced in October 2021. Like other "Mega-projects," the Northern Metropolis policy seeks to catalyze economic development in underutilized areas by addressing a broad range of issues: tackling disparities pertaining to certain locations, increasing the supply of housing, strengthening transportation connections, fostering innovation hubs and clusters, preserving natural spaces, highlighting cultural heritage, and involving local communities in decision-making processes.

Our findings reveal rapid and significant changes in human behavior shortly after the policy's announcement. Within merely four months, households responded swiftly to the policy announcement, adjusting their consumption and home-purchasing decisions, well ahead of any physical changes that were in place. This pattern is consistent with PIH, which posits that households make decisions based not only on their current income level but also on their expectations of lifetime resources. In this context, households appear to have updated their beliefs about the long-term growth potential associated with the Northern Metropolis and responded accordingly. Housing consumption thus emerges as an early indicator, often preceding tangible capital deployment and labor market changes. These findings also highlight the salience of policy narratives in shaping expectations and economic behavior well in advance of implementation.

On the firm side, the announcement triggered a rapid increase in firm establishments within the Northern Metropolis. The number of registered establishments in the I&C and construction industries rose by 14.4% and 25.7%, respectively. These findings are robust to the use of the SDiD estimator. This rapid firm-level response suggests that the policy shifted forward-looking expectations not only among households but also among firms, consistent with a "Big Push" theory in which coordinated policy signals attract complementary entry in targeted sectors.

Importantly, the effects of the policy are not uniformly distributed across the population. We document meaningful heterogeneity in responses across demographic groups. Specifically, affluent, highly educated, and family-oriented households were more likely to move into the Northern Metropolis area. These groups tend to have stronger labor market attachment and greater financial flexibility to purchase a home. These groups appear to have internalized the prospective gains from increased spatial connectedness, particularly the planned integration of Shenzhen's high-tech sector with Hong Kong's financial and professional services (e.g., an "industry-led, infrastructure-first" development approach). In contrast, disadvantaged groups such as smaller households, low-income renters, and less-educated individuals may face rising housing and living costs without proportional improvements in access to jobs or social mobility. Without complementary housing affordability and labor market policy interventions, these potential disparities may exacerbate existing inequalities in opportunity over time.

More broadly, the Northern Metropolis initiative should be viewed not in a vacuum but as an integral component of the GBA development strategy. These cross-border industrial and institutional linkages, if realized, are expected to enhance urban vibrancy by improving the flow of talent, capital, and ideas. Rising housing prices and increased consumption may serve as harbingers of the long-term reorganization of the region's economic geography.

Nonetheless, implementing these policy goals also entails considerable risks. As articulated by Flyvbjerg (2017), global spending on mega-projects constitutes approximately 8 percent of the world's GDP, yet many of these projects fall short of their intended objectives. A key concern is that efforts to reduce disparities between regions (inter-regional inequality) may unintentionally increase intra-regional disparities. For instance, investments in high-skilled sectors may raise the skill premium, resulting in wage and housing gaps between skilled and unskilled workers within the same region (Lopez and Morita, 2023). In terms of education, some neighborhoods may benefit from inflows and improved peer quality, while others may lose students and see deteriorating peer quality, potentially compromising or even offsetting overall benefits for children (Weiwu, 2025). The impact of new constructions on the value of existing housing stock also remains contested. These challenges call for targeted interventions that strike a balance between economic efficiency and social equity.

Taken together, our findings highlight the importance of recognizing both the behavioral consequences of large-scale place-based policies on individuals and households. They indicate the need for coordinated, inclusive policies that leverage the benefits of spatial transformation while mitigating its distributional risks. Hong Kong provides a valuable social laboratory to test the impact of government-led megaprojects on socioeconomic changes, but its success will ultimately hinge on how well these trade-offs are understood and managed.

Our findings can guide the development of place-based policies for other cities seeking to address regional inequality and boost economic opportunities. Our results are specifically insightful for cities situated within larger geographic/economic zones (e.g., European Union, San Francisco Bay Area, Greater Bay Area) or those undergoing structural evolution by embracing IT innovations and drawing in skilled labor (e.g., Mumbai, Kigali, Dublin).

Looking forward, our analysis opens several venues for future research. First, while our findings orbit around immediate household responses, it remains important to assess how the policy impacts firm behavior, particularly through channels such as increased collateral values and relaxed borrowing constraints (LaPoint and Sakabe, 2021). Second, the role of human capital accumulation in shaping regional development merits further attention. Skilled workers are often the first to migrate in response to new opportunities, and their presence can generate broader spillovers, either by stimulating employment growth in the non-traded sector through the "human capital multiplier" (Moretti and Thulin, 2013), or by exacerbating intra-regional inequality if skill-based gains are unevenly distributed. Third, beyond the economic well-being of households, future studies could consider the policy's implications for non-pecuniary aspects of welfare such as health, well-being, and community cohesion.

**References**

Agrawal, D. R., Hoyt, W. H., & Wilson, J. D. (2022). Local policy choice: theory and empirics. *Journal of Economic Literature*, *60*(4), 1378-1455.

Agarwal, S., Fan, Y., Qian, W. and Sing, T.F. (2026). Affordable Public Housing and Intergenerational Mobility. *Nature Cities*.

Anderson, J. E. (2011). The Gravity Model. *Annual Review of Economics*, 3(1), 133–160.

Arkhangelsky, D., Athey, S., Hirshberg, D. A., Imbens, G. W., & Wager, S. (2021). Synthetic difference-in-differences. *American Economic Review*, 111(12), 4088-4118.

Austin, B., Glaeser, E., & Summers, L. (2018). Jobs for the Heartland: Place-Based Policies in 21st-Century America. *Brookings Papers on Economic Activity*, 2018(1), 151–232.

Basheer, M., Nechifor, V., Calzadilla, A., Siddig, K., Etichia, M., Whittington, D., Hulme, D., and Harou, J. (2021). Collaborative Management of the Grand Ethiopian Renaissance Dam Increases Economic Benefits and Resilience. *Nature Communications* 12 (5622).

Bilal, A., & Rossi-Hansberg, E. (2021). Location as an Asset. *Econometrica*, *89*(5), 2459–2495.

Black, S. E., Devereux, P. J., Lundborg, P., and Majlesi, K. (2020). Poor Little Rich Kids? the Role of Nature versus Nurture in Wealth and Other Economic Outcomes and Behaviors. *Review of Economic Studies* 87(4): 1683-1725.

Cameron, C., & Miller, D. (2015). A practitioner's guide to clustered errors. *Journal of Human Resources 50*(2), 317-372.

Chen, H., Cheng, K., & Zhang, M. (2023). Does geographic proximity affect firms' cross-regional development? Evidence from high-speed rail construction in China. *Economic Modelling, 126*, 106402.

Chen, J., Glaeser, E., and Wessel, D. (2023). JUE Insight: The (non-)Effect of Opportunity Zones on Housing Prices. *Journal of Urban Economics* 133, 103451.

Cheshire, P., C.A.L. Hilber and F. Carozzi (2018) 'Measuring Affordability: Alternative Perspectives', Introduction to 14th Annual Demographia International Housing Affordability Survey.

Cobb, C. W., and Douglas, P. H. (1928). A Theory of Production. *American Economic Review*, 18(1), 139–165.

Corinth, K., & Feldman, N. (2024). Are Opportunity Zones an Effective Place-Based Policy? *Journal of Economic Perspectives*, 38(3), 113-136.

Daysal, N. M., Lovenheim, M. F., & Wasser, D. N. (2023). The Intergenerational Transmission of Housing Wealth. *National Bureau of Economic Research Working Paper* (No. w31669).

Diao, M., Kong, H., and Zhao, J. (2021). Impacts of Transportation Network Companies on Urban Mobility. *Nature Sustainability* 4, 494–500.

Ding, Y., Wang, X., & Chen, Q. (2025). Windfall gains and household consumption: Regression-discontinuity evidence from urban China's preferential housing policies. *Journal of Economic Behavior & Organization*, *240*, 107290.

Duranton, G., & Venables, A. J. (2018). Place-Based Policies for Development. NBER Working Paper Series, 24562.

Fajgelbaum, P. D., & Gaubert, C. (2025). Place-Based Policies: Lessons from Theory (No. w33517). National Bureau of Economic Research.

Flyvbjerg, B. (Ed.). (2017). *The Oxford Handbook of Megaproject Management*. Oxford University Press.

Freedman, M., & Neumark, D. (2024). Lessons Learned and Ignored in U.S. Place-Based Policymaking. NBER Working Paper Series.

Galelli, S., Dang, T., Ng, J., Chowdhury, K., and Arias, M. (2022). Opportunities to Curb Hydrological Alterations via Dam Re-operation in the Mekong. *Nature Sustainability* 5, 1058–1069.

Garmaise, M. J., Levi, Y., & Lustig, H. (2024). Spending Less after (Seemingly) Bad News. *Journal of Finance (New York)*, *79*(4), 2429–2471.

Gerritse, M., Wang, Z., & van Oort, F. (2026). Industrial Transfer Policy in China: Migration and Development. *Journal of Urban Economics*, *151*, 103815.

Glaeser, E. L., & Gottlieb, J. D. (2008). The Economics of Place-Making Policies. *Brookings Papers on Economic Activity*, *2008*(1), 155–239.

Glaeser, E. L., and Gottlieb, J. D. (2009). The Wealth of Cities: Agglomeration Economies and Spatial Equilibrium in the United States. *Journal of Economic Literature*, 47(4), 983–1028.

Glaeser, E. L., Saiz, A., Burtless, G., & Strange, W. C. (2004). The Rise of the Skilled City [with Comments]. Brookings-Wharton Papers on Urban Affairs, 2004(1), 47–105.

Holtgrieve, G. and Arias, M. (2022). Optimizing Amazonian Dams for Nature. *Science* 375 (6582): 714-715.

Jiang, H., Liang, Y., Qin, Y., & Tao, H. (2025). Place-Based Policies, Agglomeration, and Firm Innovation: Evidence from China. *Working Paper.*

Kline, P., & Moretti, E. (2014). Local Economic Development, Agglomeration Economies, and the Big Push: 100 Years of Evidence from the Tennessee Valley Authority. *Quarterly Journal of Economics*, 129(1), 275–331.

Koster, H. and Ommeren, J. (2019). Place-Based Policies and the Housing Market. *Review of Economics and Statistics* 101(3): 400-414.

LaPoint, C., and Sakabe, S. (2021). Place-based policies and the geography of corporate investment. *Available at SSRN 3950548*.

Lin, Y., Qin, Y., Wu, J., and Xu, M. (2021). Impact of High-Speed Rail on Road Traffic and Greenhouse Gas Emissions. *Nature Climate Change* 11, 952–957.

Lopez, J. C., and Morita, T. (2023). Inter-and intraregional inequality in a spatial economy. *Journal of Regional Science*.

Lu, Y., Wang, J. and Zhu, L. (2019). Place-Based Policies, Creation, and Agglomeration Economies: Evidence from China's Economic Zone Program. *American Economic Journal: Economic Policy*, 11 (3): 325-60.

Marlow, L., Lam, E., and Lindberg, K. S. (2021). Hong Kong Plans Massive City on Chinese Border as Cure for Unrest. Bloomberg.

Moran, E. F., Lopez, M. C., Moore, N., Muller, N., and Hyndman, D. W. (2018). Sustainable Hydropower in the 21st Century. *Proceedings of the National Academy of Sciences*, 115 (47): 11891-11898.

Moretti, E., & Thulin, P. (2013). Local multipliers and human capital in the United States and Sweden. *Industrial and Corporate Change*, *22*(1), 339–362.

Neumark, D., and Simpson, H. (2015). *Place-Based Policies.* Handbook of Regional and Urban Economics, 5: 1197-1287.

Rooney, R., Robinson, D. and Petrone, R. (2015). Megaproject reclamation and climate change. *Nature Climate Change* 5, 963–966.

Vilela, T., Harb, A. M., Bruner, A., Arruda, V., Ribeiro, V., Alencar, A., Grandez, A., Rojas, A., Launa, A., and Botero, R. (2020). A better Amazon Road Network for People and the Environment. *Proceedings of the National Academy of Sciences*, 117 (13): 7095-7102.

Weiwu, L. (2025). Opportunity in Motion: Equilibrium Effects of a Place-Based Policy on Economic Mobility. *Working paper*.

Yeung, P. (2022). Hong Kong Struggles to Tame a Long-Simmering Housing Crisis. Bloomberg.

Yu, Y., Fan, Y., and Yi, J (2023). One-Child Policy, Differential Fertility, and Intergenerational Transmission of Inequality in China. *NUS Working Paper*.

**Figure 1: The Impact of Northern Metropolis Policy on Housing Price**

This figure depicts event time coefficients, showcasing the differential impacts of the policy on residential properties. The coefficient for the month immediately before the policy implementation is normalized to zero. Year-month and district fixed effects are included. Standard errors are adjusted for clustering at year-month and district levels.

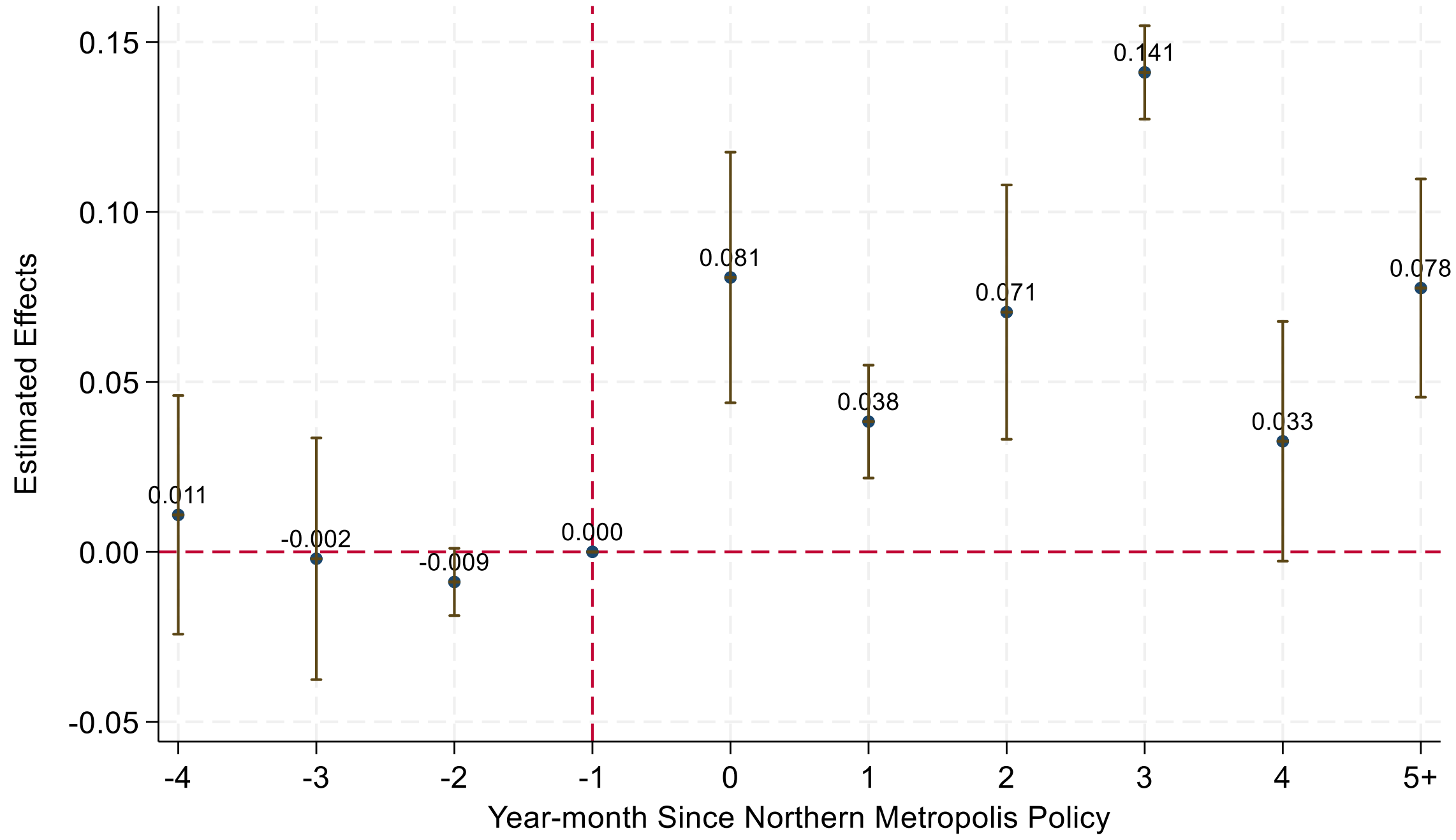

**Figure 2: The Impact of Northern Metropolis Policy on Average Amount of Online Shopping**

This figure shows the average of log online shopping order amounts placed by buyers residing within the Northern Metropolis area (treatment group) and the Non-northern Metropolis area (control group). The solid line represents the values for the Northern Metropolis area, while the dashed line represents the values for the Non-northern Metropolis area.

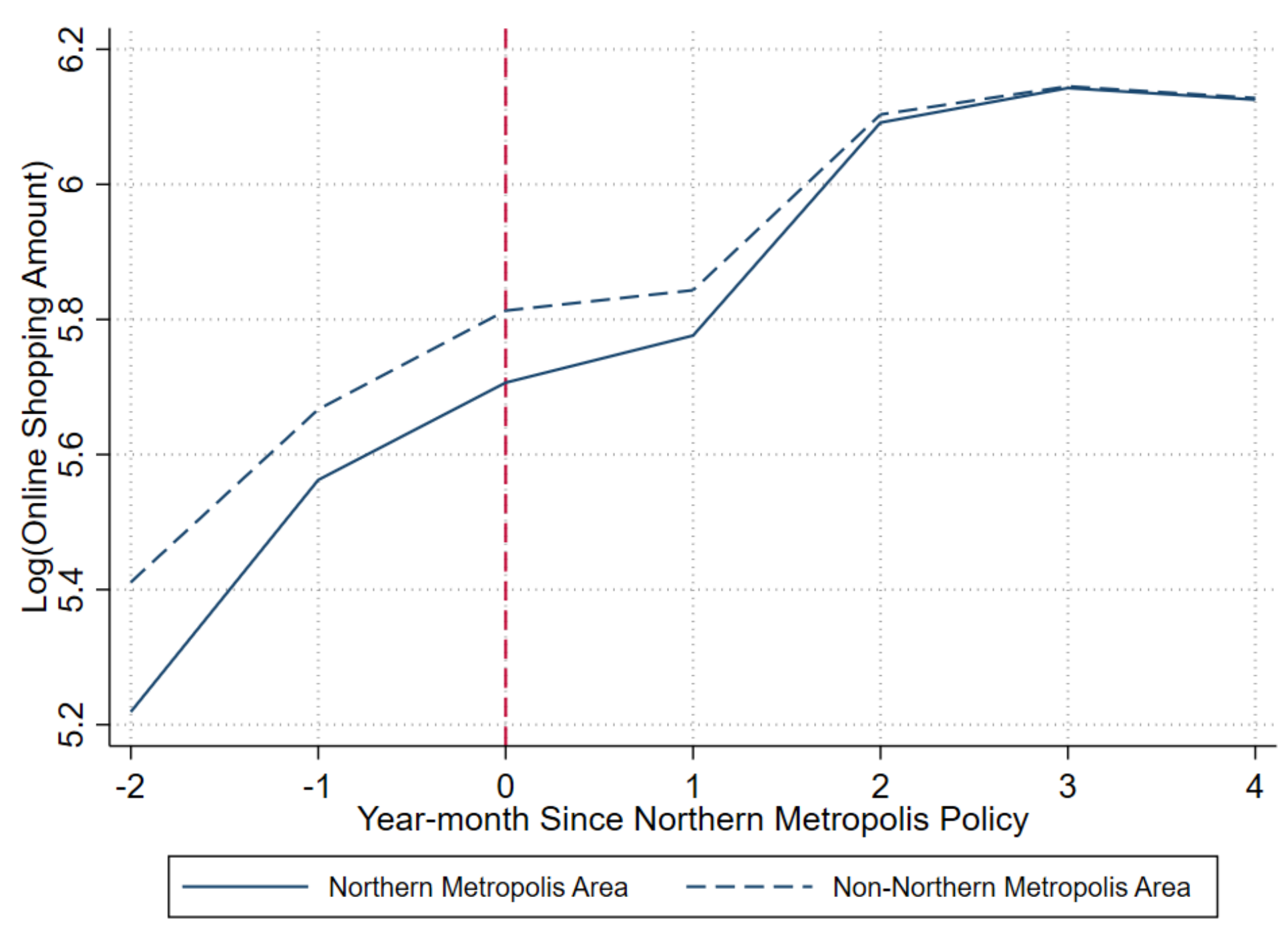

### Figure 3: The Impact of Northern Metropolis Policy on Firm Establishment

This figure depicts event time coefficients, showcasing the differential impacts of the policy on firm establishment. Panel A shows the event study figure for information and communication industry, while Panel B shows the event study figure for construction industry. The coefficient for the month immediately before the policy implementation is normalized to zero. Year-quarter and district fixed effects are included. Standard errors are adjusted for clustering at district levels.

### Panel A: Information and Communication Industry

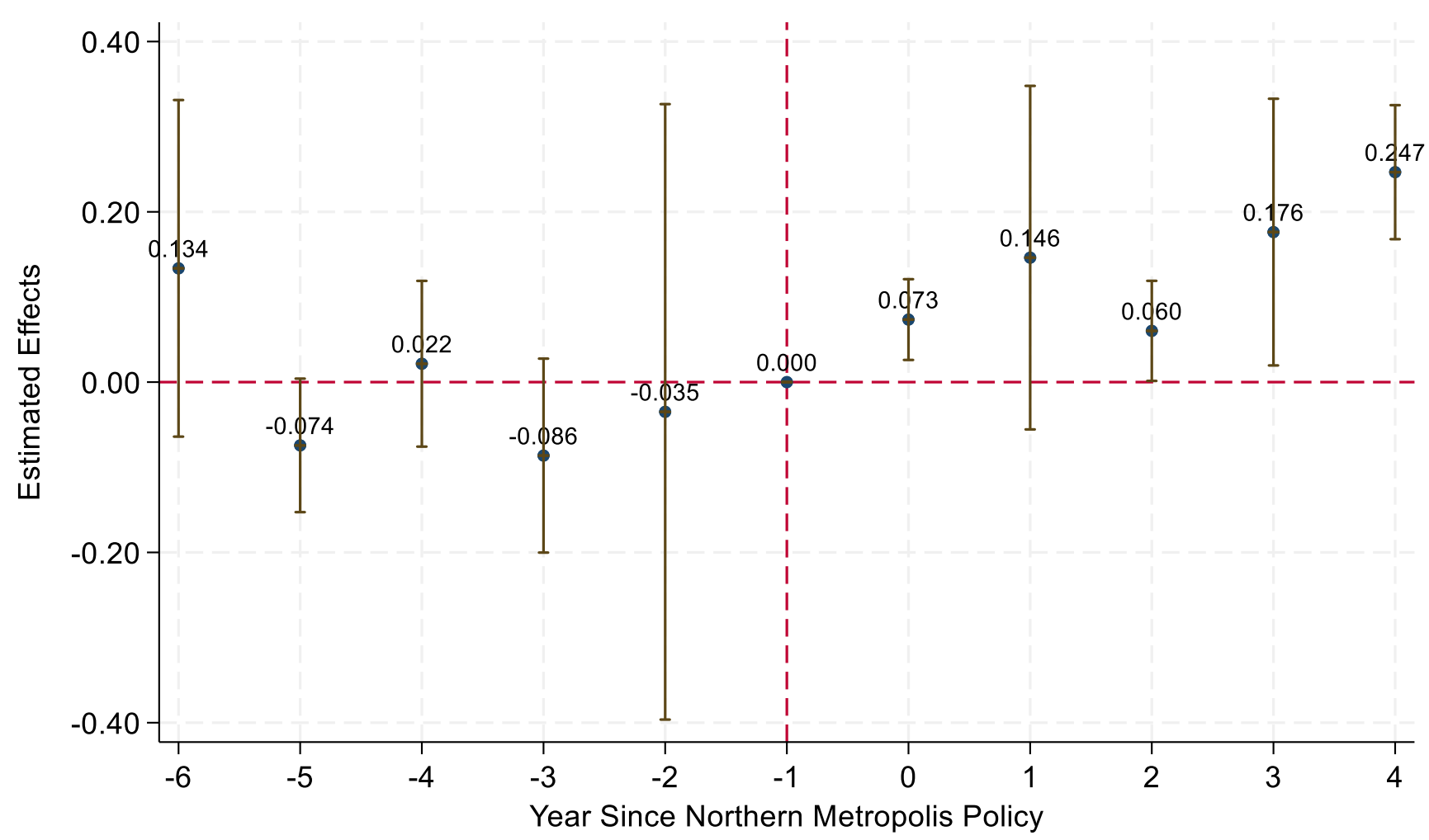

### Panel B: Construction Industry

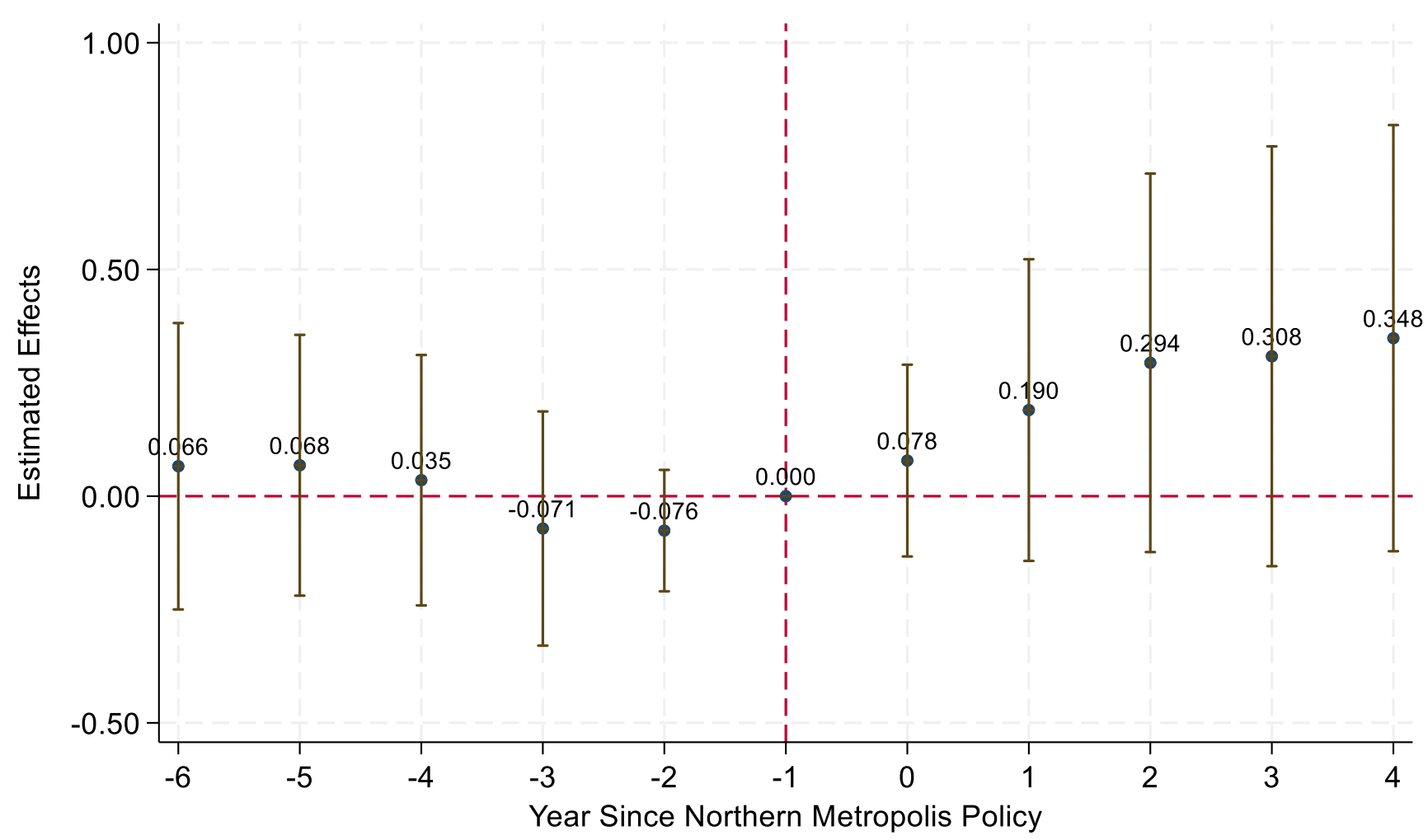

**Figure 4: The Impact of Northern Metropolis Policy on Demographic Changes**

This figure depicts event time coefficients, showcasing the differential impacts of the policy impact on higher-education residents, higher-income households, households with Children. The coefficient for the year immediately before the policy implementation is normalized to zero. Year and district fixed effects are included. Standard errors are adjusted for clustering at the district levels.

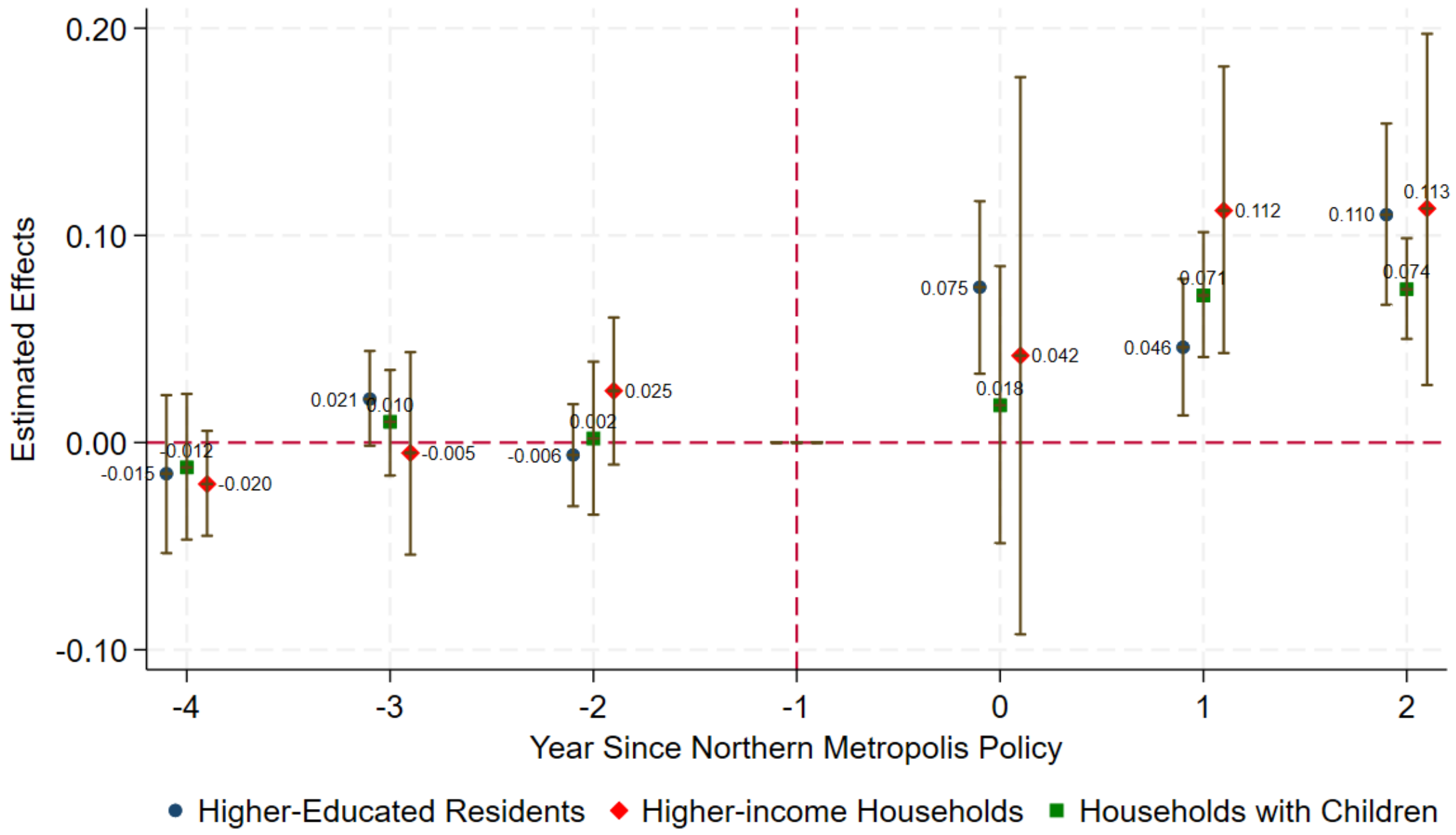

**Figure 5: The Impact of Northern Metropolis Policy on Housing Price in Bordering Area in Shenzhen Housing Market**

This figure depicts event time coefficients, showcasing the differential impacts of the policy impact on housing prices in Shenzhen housing market. The coefficient for the month immediately before the policy implementation is normalized to zero. Year-month and subdistrict fixed effects are included. Standard errors are adjusted for clustering at the year-month and district levels.

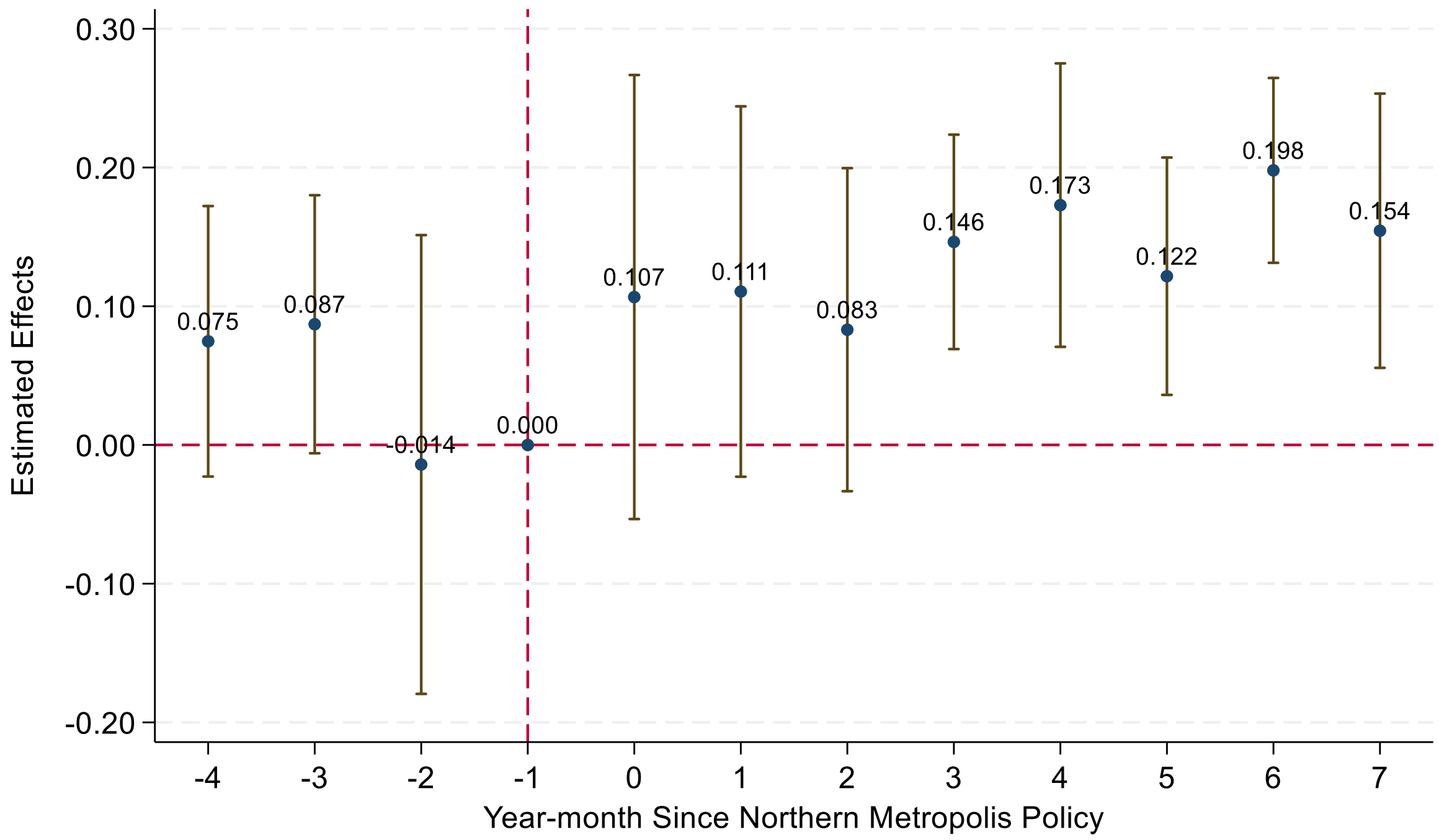

**Table 1: Policy Impact on Increasing Housing Prices in Northern Metropolis Area**

This table presents the policy impact on housing prices in the Northern Metropolis area. $Log(HomePrice)$ is defined as the logarithm of the housing transaction price in million HKD. $Policy$ indicates transactions post-October 2021 (i.e., after the announcement). $North$ is also a dichotomous variable that equals 1 if the property is located within the Northern Metropolis area and 0 otherwise. We also control property characteristics: building age, floor, and unit size. In Columns (1) and (2), results include year-month and district fixed effects. In Columns (3) and (4), we also include year-month-by-region fixed effects. Standard errors are adjusted for clustering at year-month and district levels, are included in parentheses, and ***, **, and * indicate 1%, 5%, and 10% significance, respectively.

| | (1) | (2) | (3) | (4) |
|---|---|---|---|---|
| | Y: Log(Housing Price) | | | |
| Variable | [-4, +4] Months | [-12 +12] Months | [-4, +4] Months | [-12 +12] Months |
| Policy*North | 0.0764*** | 0.0495*** | 0.0686*** | 0.0393*** |
| | (0.0225) | (0.0143) | (0.0242) | (0.0148) |
| Log(Building Age) | -0.1512*** | -0.1352*** | -0.1516*** | -0.1362*** |
| | (0.0051) | (0.0037) | (0.0050) | (0.0037) |
| Log(Floor) | 0.0740*** | 0.0749*** | 0.0748*** | 0.0754*** |
| | (0.0048) | (0.0029) | (0.0047) | (0.0028) |
| Log(Size) | 0.9994*** | 0.9803*** | 1.0001*** | 0.9798*** |
| | (0.0146) | (0.0094) | (0.0144) | (0.0093) |
| Year-month FE | YES | YES | YES | YES |
| District FE | YES | YES | YES | YES |
| Year-month*Region FE | NO | NO | YES | YES |
| Observations | 38,214 | 100,576 | 38,214 | 100,576 |
| R-squared | 0.7164 | 0.7145 | 0.7168 | 0.7153 |

**Table 2: Policy Impact on Increasing Order-level Online Shopping Amount in Northern Metropolis Area**

This table presents the policy impact on order-level online shopping amount. $Log(Order-level\ Online\ Shopping\ Amount)$ is defined as the logarithm of the order-level total shopping amount in HKD. $Policy$ is a dummy variable that equals 1 if the date of purchase is after October 2021 (i.e., after the announcement), and 0 otherwise. $North$ is also a dummy variable that equals 1 if the buyer is in the Northern Metropolis area and 0 otherwise. In addition to using full samples, we also use private housing and public housing samples separately. In Columns (1), (2), and (3), results include year-month, district, and customer fixed effects. In Columns (4), (5), and (6), we add the year-month-by-region fixed effect. Standard errors are adjusted for clustering at year-month, district, and customer levels, are included in parentheses, and ***, **, and * indicate 1%, 5%, and 10% significance, respectively.

| | (1) | (2) | (3) | (4) | (5) | (6) |
|---|---|---|---|---|---|---|
| | Y: Log(Order-level Online Shopping Amount) | | | | | |
| Variable | [-4, +4] Months | [-4, +4] Months | [-4, +4] Months | [-4, +4] Months | [-4, +4] Months | [-4, +4] Months |
| | Full samples | Private housing | Public housing | Full samples | Private housing | Public housing |
| Policy*North | 0.0678*** | 0.0788*** | 0.0331*** | 0.0412*** | 0.0469*** | 0.0260*** |
| | (0.0033) | (0.0040) | (0.0057) | (0.0034) | (0.0043) | (0.0061) |
| Year-month FE | YES | YES | YES | YES | YES | YES |
| District FE | YES | YES | YES | YES | YES | YES |
| Customer FE | YES | YES | YES | YES | YES | YES |
| Year-month*Region FE | NO | NO | NO | YES | YES | YES |
| Observations | 4,686,504 | 3,276,863 | 1,409,641 | 4,686,504 | 3,276,863 | 1,409,641 |
| R-squared | 0.4097 | 0.3963 | 0.4535 | 0.4100 | 0.3967 | 0.4536 |

**Table 3: Policy Impact on Firm Activity in Northern Metropolis Area**

This table presents the policy impact on firm establishments and employment. $Log(Establishment)$ is defined as the logarithm of the number of establishments in quarter $t$ and district $l$. $Log(Employee)$ is defined as the logarithm of the number of employees in quarter $t$ and district $l$. Column (1) and Column (2) show the results in information and communications industry, while Column (3) and Column (4) show the results in construction industry. $Policy$ is a dummy variable that equals 1 if all observations in the post-policy period (the fourth quarter of 2021 and later), and 0 otherwise. $North$ is also a dummy variable that equals 1 if the district is located in the Northern Metropolis area and 0 otherwise. We have year-quarter and district fixed effects. Standard errors are adjusted for clustering at district levels, are included in parentheses, and ***, **, and * indicate 1%, 5%, and 10% significance, respectively.

| | (1) | (2) | (3) | (4) |
|---|---|---|---|---|
| | Information and Communications Industry | | Construction Industry | |
| | Y: Log (Establishment) | Y: Log (Employee) | Y: Log (Establishment) | Y: Log (Employee) |
| Policy*North | 0.1442** | 0.1173** | 0.2572** | 0.3557** |
| | (0.0588) | (0.0497) | (0.1144) | (0.1532) |
| Year-quarter FE | YES | YES | YES | YES |
| District FE | YES | YES | YES | YES |
| Observations | 738 | 738 | 738 | 738 |
| R-squared | 0.9528 | 0.9619 | 0.8805 | 0.7271 |

**Table 4: Policy Impact on Demographic Structure Changes in Northern Metropolis Area**

This table presents the policy impact on demographic structure changes in Northern Metropolis area. The dependent variable is defined as the logarithm of annual district-level *Ratio_University*, *Higher-income households* and *Households with Larger Family Sizes* separately. *Ratio_University* is the ratio of population aged 15 holding a bachelor's degree or above. *Higher-income households* is defined as numbers of households with a monthly income exceeding HKD 30,000. *Households with Larger Family Sizes* is defined as numbers of households with two or more family members. $Policy$ is a dummy variable that equals 1 for all observations in the post-policy period (2021 and later), and 0 otherwise. $North$ is also a dummy variable that equals 1 if the data is of the Northern Metropolis area and 0 otherwise. $Outskirt$ is a dummy variable that equals 1 if the data is of the New Territories but not within the Northern Metropolis area and 0 otherwise. We also control district-level median monthly household income as macroeconomic controls. Results include year and district fixed effects. Standard errors are adjusted for clustering at the district level, are included in parentheses, and ***, **, and * indicate 1%, 5%, and 10% significance, respectively.

| | (1) | (2) | (3) | (4) | (5) | (6) |
|---|---|---|---|---|---|---|
| | Y: Log (Ratio_University) | | Y: Log (Higher-income households) | | Y: Log (Households with Larger Family Sizes) | |
| Policy*North | 0.0654*** | 0.0825*** | 0.0904** | 0.1301*** | 0.0522*** | 0.0796*** |
| | (0.0070) | (0.0062) | (0.0357) | (0.0383) | (0.0136) | (0.0151) |
| Policy*Outskirt | | 0.0398*** | | 0.0907*** | | 0.0636*** |
| | | (0.0108) | | (0.0276) | | (0.0196) |
| HHIncome | YES | YES | NO | NO | YES | YES |
| Year FE | YES | YES | YES | YES | YES | YES |
| District FE | YES | YES | YES | YES | YES | YES |
| Observations | 126 | 126 | 126 | 126 | 126 | 126 |
| R-squared | 0.9932 | 0.9942 | 0.9860 | 0.9897 | 0.9956 | 0.9967 |

**Table 5: Policy Impact on Increasing Housing Prices in Bordering Area in Shenzhen Housing Market**

This table presents the policy impact on housing prices in bordering area in Shenzhen housing market. $Log(HomePrice)$ is defined as the logarithm of the housing transaction price in million CNY. $Log(HomePrice)$ is winsorized at the 5% level. $Policy$ indicates transactions post-October 2021 (i.e., after the announcement). $Border$ is also a dichotomous variable that equals 1 if the property is located within the bordering area in Shenzhen and 0 otherwise. We define the bordering area in Shenzhen as the five border-adjacent subdistricts in Shenzhen (Yuehai, Nantou, Futian, Fubao, and Huangbei). We also control property characteristics: building age, floor, and unit size. In Columns (1) and (2), results include year-month and subdistrict fixed effects. In Columns (3) and (4), we also include year-month-by-district fixed effect. Standard errors are adjusted for clustering at year-month and district levels, are included in parentheses, and ***, **, and * indicate 1%, 5%, and 10% significance, respectively.

| | (1) | (2) | (3) | (4) |
|---|---|---|---|---|
| | Y: Log(Housing Price) | | | |
| Variable | [-4, +4] Months | [-12 +12] Months | [-4, +4] Months | [-12 +12] Months |
| Policy*Border | 0.0699* | 0.0406** | 0.0559 | 0.0368* |
| | (0.0390) | (0.0199) | (0.0418) | (0.0211) |
| Log(Building Age) | -0.1322*** | -0.0879*** | -0.1302*** | -0.0883*** |
| | (0.0272) | (0.0151) | (0.0272) | (0.0150) |
| Log(Floor) | 0.0314 | 0.0116 | 0.0332 | 0.0119 |
| | (0.0227) | (0.0080) | (0.0232) | (0.0081) |
| Log(Size) | 0.3782*** | 0.4587*** | 0.3803*** | 0.4587*** |
| | (0.0315) | (0.0212) | (0.0317) | (0.0213) |
| Year-month FE | YES | YES | YES | YES |
| Subdistrict FE | YES | YES | YES | YES |
| Year-month*District FE | NO | NO | YES | YES |
| Observations | 3,916 | 16,602 | 3,916 | 16,602 |
| R-squared | 0.2072 | 0.3248 | 0.2132 | 0.3291 |

## Appendix 1: Details of the Northern Metropolis Policy

The chart below summarizes the details of the Northern Metropolis policy.

| | Northern Metropolis Policy |
|---|---|
| Purpose of the Policies | Promote economic development in the northern part of Hong Kong and strengthen the integration of the Greater Bay area. |
| Implementation Date | October 6th, 2021 |
| Scope of the Policies | North district and Yuen Long district |
| Measures | 1) Transportation Development: The Hong Kong government will construct five new railways:<br>i. Hong Kong-Shenzhen western rail to upgrade the transport linkage with the Qianhai area in Shenzhen.<br>ii. The northern link spur line to link the Hong Kong-Shenzhen Innovation and Technology Park and new Huanggang port.<br>iii. The northern link eastward extension to Man Kam To area.<br>iv. Luohu South station, located between the existing Luohu station and Sheung Shui station on the East Rail Line.<br>v. An automated people mover system from Tsim Bei Tsui to Pak Nai.<br>2) Innovation and Technology Development: The Hong Kong government will increase land supply for innovation and technology uses in the Northern Metropolis and build the San Tin Technopole. The San Tin Technopole has an estimated area of about 1,100 hectares and can provide a total of 45,500 to 47,500 residential units. Some of the units can be used as talent apartments for I&T enterprises and research institutes.<br>3) New comprehensive development node: The Hong Kong government will increase the housing supply in the Northern Metropolis area to mitigate the housing affordability issue in Hong Kong. In the western part of the Northern Metropolis, the Hong Kong government will develop the Lau Fau Shan and Tsim Bei Tsui area, providing 47,000 to 52,500 additional residential units in the Hung Shui Kiu/Ha Tsuen New development area. In the east part of the Northern Metropolis, the Hong Kong government will develop the Lo Wu/Man Kam To, providing around 75,000 to 85,000 |
| | residential units in total, and develop the Ma Tso Lung area, providing an additional 12,000 to 13,500 residential units.<br>4) Increased environmental capacity: The Hong Kong government will resume a lot of private wetlands and fishponds in the Northern Metropolis to increase environmental capacity. In addition, the Hong Kong government will build new wetland conservation parks and expand the old wetland park in the Northern Metropolis. Hong Kong government will also develop the urban-rural greenway in the New Territories North. |

**Appendix 2: Variable Definitions**

**Panel A: Housing Transaction Data**

| **Variable Name** | **Definition** |
|---|---|
| Aggregated Transaction Amount | The building-month level transaction amount. |
| Border | A dichotomous variable that equals 1 if the property is located within the bordering area in Shenzhen and 0 otherwise. |
| Building Age | Building age in years. |
| Floor | Floor. |
| Housing Price | Property transaction price. |
| North | A dichotomous variable that equals 1 if the property is located within the Northern Metropolis area and 0 otherwise. |
| Policy | A dummy variable that indicates transactions post-October 2021 (i.e., after the announcement). |
| Size | Property saleable area in square feet. |

**Panel B: Order-level Online Consumption Data**

| Variable Name | Definition |
|---|---|
| Entertainment Shopping Amount | Customers' entertainment shopping amount in HKD per order. Entertainment shopping amount includes spending on travel, sports, toys and pets. |
| Necessity Shopping Amount | Customers' necessity shopping amount in HKD per order. Necessity shopping amount includes spending on housewares, gadgets, and supermarket goods. |
| Non-necessity Shopping Amount | Customers' non-necessity shopping amount in HKD per order. Non-necessity shopping amount includes spending on clothes, makeup, skincare, travel, sports, toys and pets. |
| Order-level Online Shopping Amount | Order-level total shopping amount in HKD. |
| Private | A dummy variable that equals 1 if the delivery address is private housing and 0 otherwise. Private residential properties include private houses, private apartments, and private village houses. |

**Panel C: Firm Activity Data**

| Variable Name | Definition |
|---|---|
| Establishment | Location-quarter level firm establishments. |
| Employee | Location-quarter level employees. |

**Panel D: Demographic Structure Data**

| Variable Name | Definition |
|---|---|
| Outskirt | A dummy variable that equals 1 if the data is of the New Territories but not within the Northern Metropolis area and 0 otherwise. |
| Ratio_University | Ratio of population aged 15 holding a bachelor's degree or above (including first degree, taught postgraduate and research postgraduate qualifications from local or non-local institutions). |
| HHIncome | District-level median monthly household income in HKD. |
| Higher-income Households | Numbers of households with a monthly income exceeding HKD 30,000 (the highest income tier in the Census classification). |
| Households with Larger Family Sizes | Number of households with two or more family members. |

## Appendix 3: Channels of Regional Spillover

Several mechanisms generate cross-border spillovers. First, spatial displacement implies zero-sum relocation. Firms or households move from nearby areas into the treated zone to capture policy benefits. The empirical prediction is a rise in activity within the policy area accompanied by stagnation or decline in adjacent locations (Neumark and Simpson, 2015; Corinth and Feldman, 2024).

Second, agglomeration spillovers (we show productivity gain in the previous subsection) imply positive-sum effects. If the policy successfully jumpstarts a cluster, productivity gains can extend beyond administrative borders through knowledge diffusion, supplier networks, and thicker labour markets. The empirical prediction is rising prices or activity both inside the treated area and in nearby locations, potentially with a lag (Fajgelbaum and Gaubert, 2025).

Third, labour market integration can transmit benefits spatially. When workers commute across boundaries, improved job prospects in the policy area may increase income and housing demand in neighbouring districts. This predicts spillovers in residential outcomes even when production remains concentrated in the treated zone (Neumark and Simpson, 2015).

Fourth, capitalization effects arise when expectations of future development are priced into nearby real estate markets. Forward-looking buyers may bid up housing prices in adjacent areas viewed as close substitutes. This predicts housing price spillovers even in the absence of immediate changes in local employment or amenities (Corinth and Feldman, 2024).

These channels jointly imply that spillovers are a central feature of large-scale place-based interventions rather than an anomaly. We therefore examine housing and firm responses in Shenzhen subdistricts bordering the Northern Metropolis to assess whether the policy triggered broader regional adjustment.

## Appendix 4: The Long-term Impact of Northern Metropolis Plan

We begin our prediction with the Cobb-Douglas production function.

$$Y_t = A_t K_t^{\alpha} L_t^{1-\alpha} \quad \text{---(2)}$$

where $Y_t$ denotes the aggregate output, $A_t$ is total factor productivity (TFP), and the factor inputs are capital ($K_t$) and labor ($L_t$). In our estimation, we proxy for the aggregate output using Hong Kong's Gross Domestic Product (GDP) in million HKD. Capital and labor are gauged by Hong Kong's gross value on investment expenditure (in million HKD) and the number of employed persons, respectively. Taking log differences leads to:

$$\Delta lnY_t = (1-\alpha)\Delta lnL_t + \alpha\Delta lnK_t + \Delta lnA_t \quad \text{---(2a)}$$

Rearranging Equation (2a), we estimate $\alpha$ from:

$$\Delta lnY_t = \Delta lnL_t + \Delta lnA_t + \alpha(\Delta lnK_t - \Delta lnL_t) \quad \text{---(2b)}$$

Based on annual data from 1997 to 2021, prior to the policy announcement, we estimated that capital elasticity, $\alpha$, is 0.350, statistically significant at the 1% level.

According to the Northern Metropolis action agenda, the area is expected to generate approximately 500,000 job opportunities, including 150,000 positions in the Information & Technology (I&T) sector. Relative to the existing level of employed persons in 2023 (3,709.6 thousand people), $\Delta lnL_t$, the log change in the labor input will be 0.126 (ln (3,709.6+500)-ln (3709.6)). We assume that total factor productivity (TFP) remains constant ($\Delta lnA_t = 0$) and the fixed investment expenditure mirrors the scale of the 'Lantau Tomorrow Vision' initiated in 2018, approximately HKD 624,000 million (USD 79,490 million).[9] Given the existing level of investment expenditure of HKD 501,457 million (USD 63,879.9 million) in 2023, $\Delta lnK_t$, the log change in the capital input will be 0.808 (ln (501,457+624,000)-ln (501,457)). Given the capital elasticity of 0.350, $\Delta lnY_t$, the log change in aggregate output will be 0.365 (0.65*0.126+0.35*0.808). Given the existing level of GDP HKD 2,901 billion (USD 369.6 billion) in 2023, we predict a projected increase in GDP to around HKD 4,180 billion (USD 532.5 billion) ($e^{\ln(2,901,018)+0.365}$), marking an increment of approximately HKD 1,279 (4,180-2,901) billion (USD 162.9 billion) attributable to the Northern Metropolis policy. This is roughly double the projected investment expenditure for the project (1,279 / 624), resulting in a 44.1% growth in GDP compared to the 2023 baseline of HKD 2,901 billion (USD 369.6 billion).

[9] https://www.devb.gov.hk/en/home/my_blog/index_id_330.html

Wages ($w_t$) follow from the marginal product of labor:

$$w_t = \frac{\partial Y_t}{\partial L_t} = (1-\alpha)A_t K_t^{\alpha} L_t^{-\alpha} \quad \text{---(3)}$$

We proxy for wages using the median monthly salary. Taking log differences gives:

$$\Delta lnw_t = \alpha(\Delta lnK_t - \Delta lnL_t) + \Delta lnA_t \quad \text{---(4)}$$

Under the assumption of constant TFP, we predict an increase in the logarithm of the average monthly salary ($\Delta lnw_t$=0.239, or 0.350 *(0.808-0.126)). Given the existing level of median monthly salary HKD 20,700 (USD 2,636.9) in 2023, we forecast a surge in the median monthly salary to reach HKD 26,289 ($e^{\ln(20,700)+0.239}$) (USD 3,348.9), amounting to an increase of HKD 5,589 (26,289-20,700) (USD 712) when the Northern Metropolis is fully operational.

Housing wealth represents an important source of wealth accumulation. Building upon Glaeser and Gottlieb's (2009) extension of the Cobb-Douglas model, we forecast the changes in housing prices after the Northern Metropolis is fully established. The pivotal idea here is the competition for land use between the tradable (industrial) and non-tradable (residential) sectors drives housing prices:

$$\Delta lnP_t = \lambda_1 \Delta lnInd_t + \lambda_2 \Delta lnRes_t + \sigma_t \quad \text{---(5)}$$

where $P_t$ represents the housing price index. $Ind_t$ measures the annual industrial new completions in square meters.[10] $Res_t$ denotes the annual private residential new completions in units. $\sigma_t$ denotes local amenities.

Using data predating the policy announcement, the estimated coefficient $\lambda_1$ is 0.055. The coefficient

[10] The categories of industrial new completions include private flatted factories, private specialized factories, and private storage facilities.

$\lambda_2$ has a value of -0.138. The value of $\sigma_t$ is 0.052. All coefficient estimates are statistically significant at least at the 10% level. Conservatively speaking, according to the action agenda for the Northern Metropolis policy, we estimate the area will create 71,940 new private residential units in major development projects.[11] Meanwhile, we estimate the area will provide 7,720,000 square meters in new industrial completions.[12]

Compared to the new completions of industrial space (113,200 square meters) in 2023, we estimate that $\Delta lnInd_t$ will be 4.222 (ln(7,720,000)-ln(113,200)) after the establishment in Northern Metropolis policy. Similarly, compared to the new completions of private housing (13,852) units in 2023, we estimate that $\Delta lnRes_t$ will be 1.647 (ln(71,940)-ln(13,852)). Assuming the $\sigma_t$ as 0.052, the change in the logarithm form of the housing price index, or $\Delta lnP_t$ will be 0.057. Given the existing level of housing price (337.400) in 2023, we predict an increase in housing price index to 357.152 ($e^{\ln(337.400)+0.057}$) and conjecture a long-term increase in housing price by approximately 5.9% (357.152 /337.400-1).

We also estimate the changes in housing prices following a pre-policy increase (i.e., the Northern Metropolis policy had not been enacted). For the residential sector, we first calculate the average year-level units of new completions using the recent ten years of data predating the Northern Metropolis policy, and the average number is 14,767 units. Assuming that new completions in the Northern Metropolis area account for 10% of total new completions, we thus predict a total of 29,535 units of new residential completions in that area over the next two decades (14,767*10*20). Similarly, for the industrial sector, we first calculate the average year-level square meters of new completions using the recent ten years of data predating the Northern Metropolis policy, and the average number is 125,120 square meters. Therefore, we estimate that the total new industrial completions in the Northern Metropolis area over the next two decades will amount to 2,502,400 square meters (125,120*20).

---

[11] Despite an initial estimate of approximately 500,000 housing units to be delivered through this policy, we have taken a conservative approach and calculated a cumulative estimated supply of 239,800 units from the major development projects. In the meantime, the ratio of public and private residential supply in Hong Kong in the vast majority of cases will be seven to three. We thus estimate the new private housing supply will be 71,940 units (239,800*0.3) in that area.

[12] According to the action agenda, the San Tin Technopole and Industrial building near Hung Shui Kiu will provide a gross floor area of 7 million square meters and 720,000 square meters, respectively.

Considering the new completions of industrial space (113,200 square meters) in 2023, we estimate that $\Delta lnInd_t$ will be 3.096 (ln(2,502,400)-ln(113,200)). Similarly, comparing the new completions of private housing (13,852 units) in 2023, we estimate that $\Delta lnRes_t$ will be 0.757 (ln(29,535)-ln(13,852)). Assuming the $\sigma_t$ as 0.052, the change in the logarithm form of price index, or $\Delta lnP_t$ will be 0.118. Given the existing level of housing prices in 2023, which is 337.400, we predict that the housing price index will increase to 379.577 ($e^{\ln(337.400)+0.118}$) and conjecture a long-term increase in housing price by approximately 12.5% if the Northern Metropolis policy had not been enacted.

In contrast, our analysis predicts a more modest housing price increase of 5.9% following the implementation of the Northern Metropolis policy. This modest rise helps maintain housing affordability and mitigates potential adverse effects on the residential market. While our calculations do not indicate significant housing price increases/declines beyond fundamentals, it is important to acknowledge that housing prices may still fluctuate due to unforeseen changes in economic fundamentals. For instance, policy uncertainty and investor speculation could introduce variability and lead to realized price changes that diverge from our predictions.

Next, to estimate the expected influx of immigrants following the establishment of the Northern Metropolis, we employ the gravity law of population migration model (Anderson, 2011), expressed as follows:

$$lnM_{ijt} = \beta_1 lnPop_{it} + \beta_2 lnPop_{jt} - \gamma lnD_{ij} + \varepsilon_{ijt} \quad \text{---(6)}$$

where $M_{ijt}$ denotes the cumulative migration flow from area $i$ to area $j$ at year level. $Pop_{it}$ and $Pop_{jt}$ represent the populations (in thousands) of the originating area $i$ and the destination area $j$, respectively. $D_{ij}$ is the shortest flight distance in kilometers between areas $i$ and $j$. We utilize the OECD international migration database, which provides country-level migration data from 2000, and the Penn World Table, which provides country-level population data up until 2019. Therefore, we limit our analysis to the period between these two years. The estimated coefficient $\beta_1$ is 0.326. The coefficient $\beta_2$ is 0.428. The coefficient $\gamma$ is 0.702. All coefficient estimates are statistically significant at least at the 1% level.

The coefficient $\beta_2$ indicates that a 1% increase in Hong Kong's population (destination area) is associated with a 0.428% increase in the cumulative migrants from other regions to Hong Kong. Given the current population level (7,490.336 thousand people) and the average annual growth rate of Hong Kong's population (0.60%) over the sample period, along with the projection of 500,000 new job opportunities from the Northern Metropolis initiative, we predict that Hong Kong's population will increase by 19.35% over the next two decades ((7,490.336*(1+0.60%)^20+500)/7,490.336-1). Therefore, we expect an 8.28% (19.35 *0.428%) increase in the cumulative migrants from other regions to Hong Kong.

However, it is important to note that our migration estimation should be interpreted as a lower bound. This is because, although the gravity model incorporates international migration data from all potential countries, the potential migration inflow from each country can vary significantly in reality. Furthermore, distance does not always have a similar effect—migration from nearby countries may be limited by restrictive geopolitics, while more distant countries may have higher flows. These factors may cause our model to underestimate the true migration potential; therefore, our predicted increase in migration should be regarded as a conservative estimate.

## Appendix 5: Supplementary Tables

### Figure A1: The Impact of Northern Metropolis Policy on Aggregated Transaction Amount

This figure depicts event time coefficients, showcasing the differential impacts of the policy impact on residential properties. The coefficient for the month immediately before the policy implementation is normalized to zero. Year-month, district and year-month-by-region fixed effects are included. Standard errors are adjusted for clustering at the district levels.

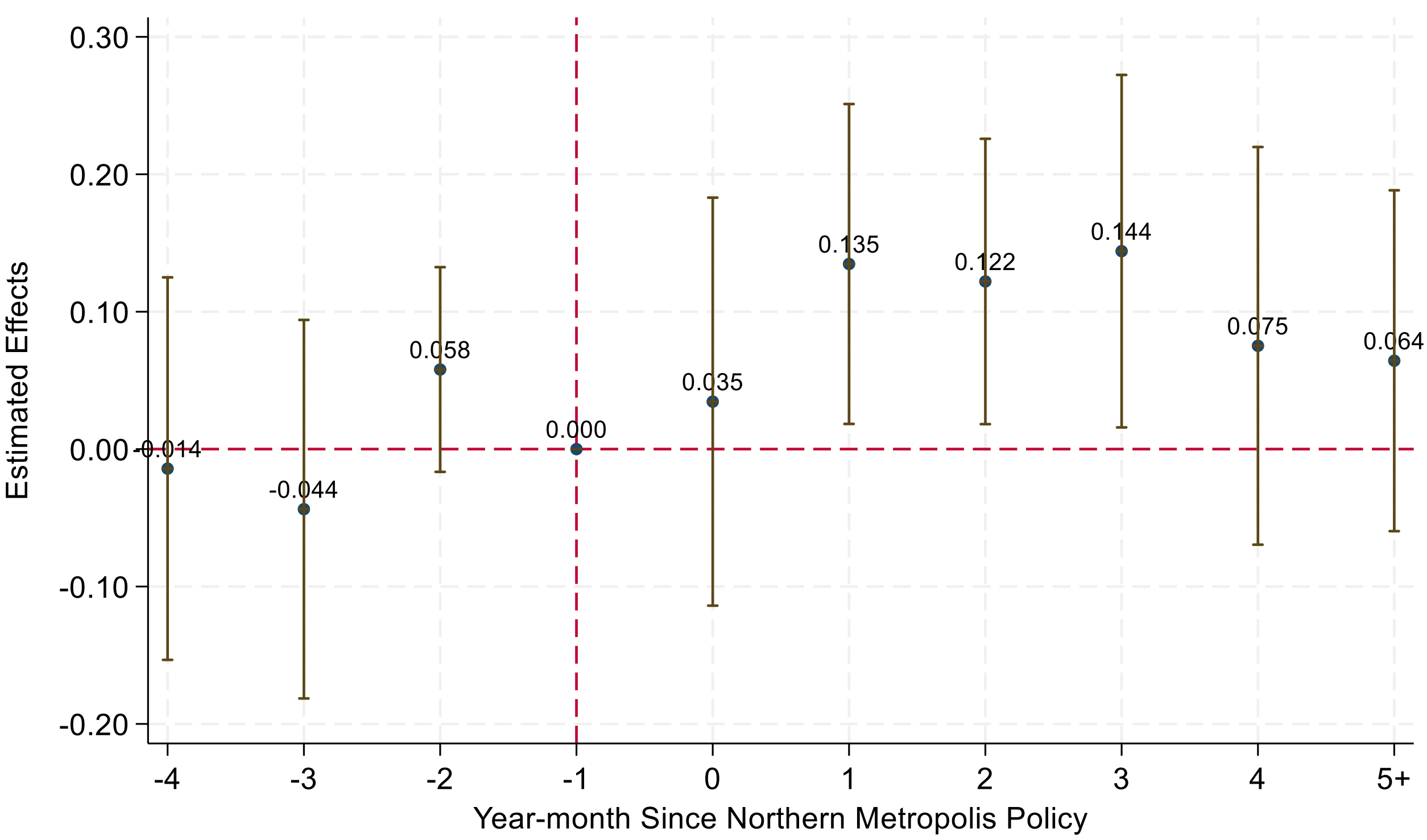

**Figure A2: The Impact of Northern Metropolis Policy on Aggregated Transaction Amount in Bordering Area in Shenzhen Housing Market**

This figure depicts event time coefficients, showcasing the differential impacts of the policy impact on transaction amount in bordering area in Shenzhen housing market. The coefficient for the month immediately before the policy implementation is normalized to zero. Year-month, subdistrict and year-month-by-district fixed effects are included. Standard errors are adjusted for clustering at the year-month and district levels.

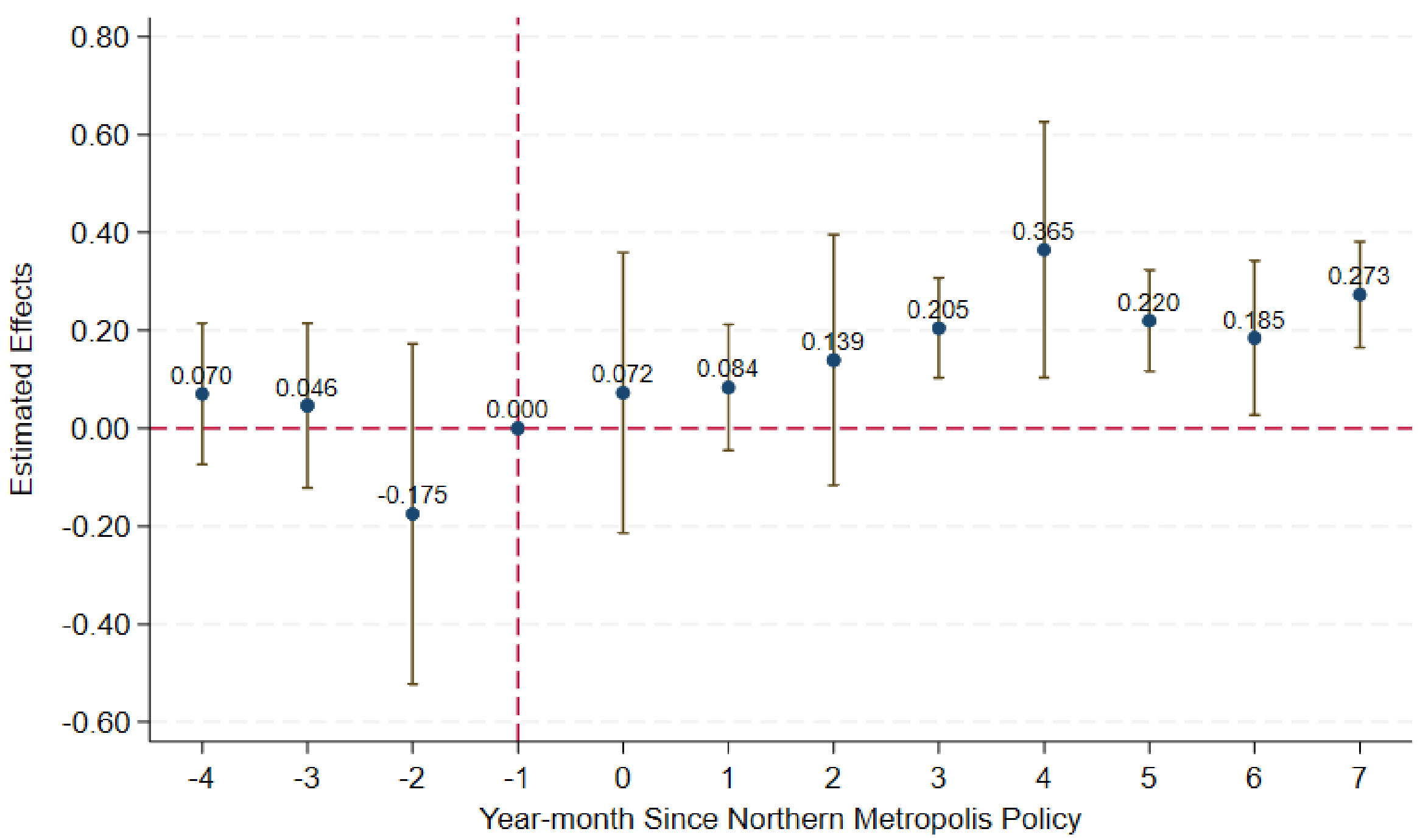

**Figure A3: The Long-term Impact of Northern Metropolis Policy**

This figure visualizes the long-term impact of the policy on four dimensions: economic development, income growth, housing wealth, and migration.

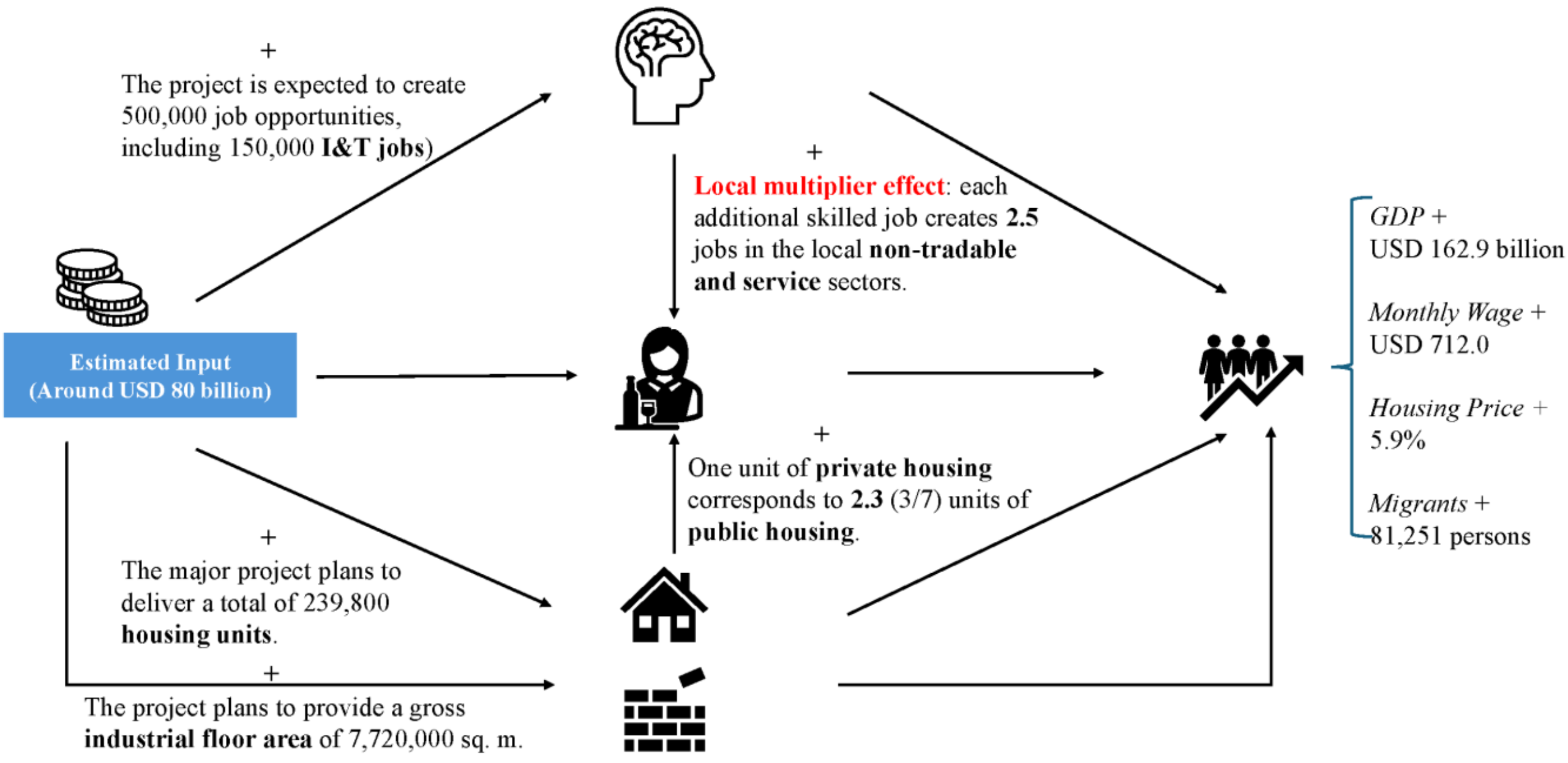

**Table A1: Summary Statistics**

This table provides summary statistics for key variables used in this study. The detailed variable descriptions are in Appendix 2.

**Panel A: Housing Transaction Data in Hong Kong**

| Variable | N | mean | SD | P25 | P50 | p75 |
|---|---|---|---|---|---|---|
| Housing Price (in millions HKD) | 100,576 | 9.331 | 10.151 | 5.726 | 7.594 | 10.020 |
| Size | 100,576 | 545.083 | 268.586 | 379.000 | 492.000 | 645.000 |
| Floor | 100,576 | 17.786 | 12.988 | 8 | 15 | 25 |
| Building Age | 100,576 | 17.892 | 15.934 | 1 | 18 | 31 |
| North | 100,576 | 0.134 | 0.341 | 0 | 0 | 0 |

**Panel B: Online Consumption Data**

| Variable | N | mean | SD | P25 | P50 | p75 |
|---|---|---|---|---|---|---|
| Order-level Online Shopping Amount | 4,686,504 | 568.750 | 539.300 | 324.500 | 474.500 | 674.500 |
| Necessity Shopping Amount | 4,686,504 | 453.644 | 474.060 | 124.500 | 374.500 | 574.500 |
| Non-necessity Shopping Amount | 4,686,504 | 115.106 | 281.919 | 0.000 | 0.000 | 107.693 |
| Entertainment Shopping Amount | 4,686,504 | 34.004 | 149.242 | 0 | 0 | 0 |
| North | 4,686,504 | 0.105 | 0.307 | 0 | 0 | 0 |
| Private | 4,686,504 | 0.699 | 0.459 | 0 | 1 | 1 |

**Panel C: Quarter-district Level Firm Establishment Data**

| Variable | N | mean | SD | P25 | P50 | p75 |
|---|---|---|---|---|---|---|
| Information & Communication Industry | | | | | | |
| Establishment | 738 | 827.203 | 592.795 | 363 | 681 | 1,180 |
| Employee | 738 | 5,996.102 | 5,882.109 | 1,730 | 4,332 | 8,574 |
| Construction Industry | | | | | | |
| Establishment | 738 | 86.963 | 35.306 | 61 | 82 | 108 |
| Employee | 738 | 6,031.970 | 3,676.191 | 3,325 | 5,371 | 7,511 |

**Panel D: Year-district Level Demographic Structure Data**

| Variable | N | mean | SD | P25 | P50 | p75 |
|---|---|---|---|---|---|---|
| Ratio_University | 126 | 27.205 | 8.435 | 18.900 | 27.150 | 31.900 |
| Higher-income Households (000) | 126 | 70.027 | 23.846 | 53.500 | 64.000 | 90.600 |
| Households with Larger Family Sizes (000) | 126 | 118.033 | 46.617 | 86.800 | 118.050 | 147.700 |

**Panel E: Housing Transaction Data in Shenzhen**

| Variable | N | mean | SD | P25 | P50 | p75 |
|---|---|---|---|---|---|---|
| Home Price (in millions CNY) | 16,602 | 4.280 | 2.252 | 2.440 | 4.088 | 6.060 |
| Size | 16,602 | 809.900 | 408.269 | 489.870 | 778.076 | 1002.667 |
| Floor | 16,602 | 12.501 | 8.765 | 5 | 11 | 19 |
| Building Age | 16,602 | 18.605 | 5.915 | 15.258 | 18.052 | 21.704 |
| Border | 16,602 | 0.261 | 0.439 | 0 | 0 | 0 |

**Panel F: Quarter-subdistrict Level Firm Establishment Data in Shenzhen**

| Variable | N | mean | SD | P25 | P50 | p75 |
|---|---|---|---|---|---|---|
| Information & Communication Industry Establishment | 2,856 | 66.189 | 82.695 | 22 | 39 | 77 |
| Construction Industry Establishment | 3,375 | 9.332 | 12.026 | 2 | 5 | 11 |

**Table A2: Policy Impact on Increasing Housing Price in Northern Metropolis Area: Robustness Check with Samples within New Territories and Winsorized Home Price**

This table presents the policy impact on housing prices in the Northern Metropolis area. In this table, we only use samples within the New Territory area in Hong Kong. $Log(HomePrice)$ is defined as the logarithm of the housing transaction price in million HKD. $Policy$ indicates transactions post-October 2021 (i.e., after the announcement). $North$ is also a dichotomous variable that equals 1 if the property is located within the Northern Metropolis area and 0 otherwise. We also control property characteristics: building age, floor, and property size. In Column (3) and Column (4), we use winsorized home prices. We include year-month, district, and year-month-by-region fixed effects. Standard errors are adjusted for clustering at year-month and district levels, are included in parentheses, and ***, **, and * indicate 1%, 5%, and 10% significance, respectively.

| | (1) | (2) | (3) | (4) |
|---|---|---|---|---|
| | Y: Log(Home Price) | | Y: Winsorized Log(Home Price) | |
| Variable | [-4, +4] Months | [-12 +12] Months | [-4, +4] Months | [-12 +12] Months |
| Policy*North | 0.0731*** | 0.0439*** | 0.0623*** | 0.0407*** |
| | (0.0246) | (0.0149) | (0.0189) | (0.0130) |
| Log(Building Age) | -0.1369*** | -0.1236*** | -0.1322*** | -0.1209*** |
| | (0.0054) | (0.0036) | (0.0048) | (0.0035) |
| Log(Floor) | 0.0530*** | 0.0534*** | 0.0581*** | 0.0545*** |
| | (0.0060) | (0.0033) | (0.0053) | (0.0030) |
| Log(Size) | 0.9096*** | 0.8780*** | 0.8703*** | 0.8580*** |
| | (0.0160) | (0.0090) | (0.0096) | (0.0070) |
| Year-month FE | YES | YES | YES | YES |
| District FE | YES | YES | YES | YES |
| Year-month*Region FE | YES | YES | YES | YES |
| Observations | 21,709 | 58,828 | 21,709 | 58,828 |
| R-squared | 0.6384 | 0.6502 | 0.7428 | 0.7436 |

**Table A3: Policy Impact on Increasing Housing Transaction Amount in Northern Metropolis Area**

This table presents the policy impact on aggregated transaction amount in the Northern Metropolis area. $Log(Aggregated\ Transaction\ Amount)$ is defined as the logarithm of the building-month level transaction amount. $Policy$ indicates transactions post-October 2021 (i.e., after the announcement). $North$ is also a dichotomous variable that equals 1 if the property is located within the Northern Metropolis area and 0 otherwise. In Columns (1) and (2), results include year-month and district fixed effects. In Columns (3) and (4), we also include year-month-by-region fixed effect. Standard errors are adjusted for clustering at district levels, are included in parentheses, and ***, **, and * indicate 1%, 5%, and 10% significance, respectively.

| | (1) | (2) | (3) | (4) |
|---|---|---|---|---|
| | Y: Log(Aggregated Transaction Amount) | | | |
| Variable | [-4, +4] Months | [-12 +12] Months | [-4, +4] Months | [-12 +12] Months |
| Policy*North | 0.0464* | 0.0794*** | 0.0822*** | 0.1164*** |
| | (0.0271) | (0.0222) | (0.0273) | (0.0211) |
| Year-month FE | YES | YES | YES | YES |
| District FE | YES | YES | YES | YES |
| Year-month*Region FE | NO | NO | YES | YES |
| Observations | 20,366 | 52,704 | 20,366 | 52,704 |
| R-squared | 0.1582 | 0.1496 | 0.1589 | 0.1507 |

**Table A4: Policy Impact on Increasing Order-level Online Shopping Amount in Northern Metropolis Area: Robustness Check with Samples within New Territories and Winsorized Shopping Amount**

This table presents the policy impact on order-level online shopping amount in the Northern Metropolis area. In this table, we only use samples within the New Territory area in Hong Kong. $Log(Order-level\ Online\ Shopping\ Amount)$ is defined as the logarithm of the order-level total shopping amount in HKD. $Policy$ is a dummy variable that equals 1 if the date of purchase is after October 2021 (i.e., after the announcement), and 0 otherwise. $North$ is also a dummy variable that equals 1 if the buyer is in the Northern Metropolis area and 0 otherwise. In Column (4), Column (5), and Column (6), we use the winsorized shopping amount. We include year-month, district, customer, and year-month-by-region fixed effects. Standard errors are adjusted for clustering at year-month, district, and customer levels, are included in parentheses, and ***, **, and * indicate 1%, 5%, and 10% significance, respectively.

| | (1) | (2) | (3) | (4) | (5) | (6) |
|---|---|---|---|---|---|---|
| | Y: Log(Order-level Online Shopping Amount) | | | Y: Winsorized Log(Order-level Online Shopping Amount) | | |
| Variable | [-4, +4] Months | [-4, +4] Months | [-4, +4] Months | [-4, +4] Months | [-4, +4] Months | [-4, +4] Months |
| | Full samples | Private housing | Public housing | Full samples | Private housing | Public housing |
| Policy*North | 0.0415*** | 0.0471*** | 0.0264*** | 0.0412*** | 0.0468*** | 0.0263*** |
| | (0.0035) | (0.0043) | (0.0061) | (0.0035) | (0.0043) | (0.0061) |
| Year-month FE | YES | YES | YES | YES | YES | YES |
| District FE | YES | YES | YES | YES | YES | YES |
| Customer FE | YES | YES | YES | YES | YES | YES |
| Year-month*Region FE | YES | YES | YES | YES | YES | YES |
| Observations | 2,474,406 | 1,657,169 | 817,237 | 2,474,406 | 1,657,169 | 817,237 |
| R-squared | 0.4208 | 0.4068 | 0.4620 | 0.4204 | 0.4063 | 0.4616 |

**Table A5: Policy Impact on Increasing Online Shopping Amount by Consumption Category in Northern Metropolis Area**

This table presents the policy impact on online shopping amount by consumption category in the Northern Metropolis area. The dependent variable is defined as the logarithm of categorized spending. Necessity shopping amount includes spending on housewares, gadgets, and supermarket goods. Non-necessity shopping amount includes spending on clothes, makeup, skincare, travel, sports, toys, and pets. Entertainment shopping amount includes spending on travel, sports, toys, and pets. $Policy$ is a dummy variable that equals 1 if the date of purchase is after October 2021 (i.e., after the announcement), and 0 otherwise. $North$ is also a dummy variable that equals 1 if the buyer is in the Northern Metropolis area and 0 otherwise. We investigate the policy impact on necessities, non-necessities, and entertainment separately. In Column (1), Column (3), and Column (5), we use private housing samples. In Column (2), Column (4), and Column (6), we use public housing samples. Results include year-month, district, customer, and year-month-by-region fixed effects. Standard errors are adjusted for clustering at year-month, district, and customer levels, are included in parentheses, and ***, **, * indicate 1%, 5%, and 10% significance, respectively.

| | (1) | (2) | (3) | (4) | (5) | (6) |
|---|---|---|---|---|---|---|
| | Y: Log(Necessity Shopping Amount) | | Y: Log(Non-necessity Shopping Amount) | | Y: Log(Entertainment Shopping Amount) | |
| Variable | [-4, +4] Months | [-4, +4] Months | [-4, +4] Months | [-4, +4] Months | [-4, +4] Months | [-4, +4] Months |
| | Private housing | Public housing | Private housing | Public housing | Private housing | Public housing |
| Policy*North | 0.0409*** | 0.0182** | 0.0453*** | 0.0113 | 0.0253*** | 0.0116 |
| | (0.0062) | (0.0089) | (0.0096) | (0.0135) | (0.0060) | (0.0082) |
| Year-month FE | YES | YES | YES | YES | YES | YES |
| District FE | YES | YES | YES | YES | YES | YES |
| Customer FE | YES | YES | YES | YES | YES | YES |
| Year-month*Region FE | YES | YES | YES | YES | YES | YES |
| Observations | 3,276,863 | 1,409,641 | 3,276,863 | 1,409,641 | 3,276,863 | 1,409,641 |
| R-squared | 0.3613 | 0.4519 | 0.3728 | 0.4497 | 0.4597 | 0.5558 |

**Table A6: Policy Impact on Increasing Online Shopping Amount by Consumption Category in Northern Metropolis Area: Robustness Check with Samples within New Territories and Winsorized Shopping Amount**

This table presents the policy impact on online shopping amount by consumption category in the Northern Metropolis area. In this table, we only use samples within the New Territory area in Hong Kong. The dependent variable is defined as the logarithm of categorized spending. The dependent variable is winsorized at 1% level. $Policy$ is a dummy variable that equals 1 if the date of purchase is after October 2021 (i.e., after the announcement), and 0 otherwise. $North$ is also a dummy variable that equals 1 if the buyer is in the Northern Metropolis area and 0 otherwise. Results include year-month, district, customer, and year-month-by-region fixed effects. Standard errors are adjusted for clustering at year-month, district, and customer levels, are included in parentheses, and ***, **, and * indicate 1%, 5%, and 10% significance, respectively.

| | (1) | (2) | (3) | (4) | (5) | (6) |
|---|---|---|---|---|---|---|
| | Y: Winsorized Log(Necessity Shopping Amount) | | Y: Winsorized Log(Non-necessity Shopping Amount) | | Y: Winsorized Log(Entertainment Shopping Amount) | |
| Variable | [-4, +4] Months | [-4, +4] Months | [-4, +4] Months | [-4, +4] Months | [-4, +4] Months | [-4, +4] Months |
| | Private housing | Public housing | Private housing | Public housing | Private housing | Public housing |
| Policy*North | 0.0411*** | 0.0197** | 0.0427*** | 0.0096 | 0.0241*** | 0.0105 |
| | (0.0062) | (0.0089) | (0.0096) | (0.0136) | (0.0059) | (0.0082) |
| Year-month FE | YES | YES | YES | YES | YES | YES |
| District FE | YES | YES | YES | YES | YES | YES |
| Customer FE | YES | YES | YES | YES | YES | YES |
| Year-month*Region FE | YES | YES | YES | YES | YES | YES |
| Observations | 1,657,169 | 817,237 | 1,657,169 | 817,237 | 1,657,169 | 817,237 |
| R-squared | 0.3739 | 0.4598 | 0.3780 | 0.4544 | 0.4674 | 0.5593 |

**Table A7: Policy Impact on Firm Activity in Northern Metropolis Area (Synthetic Difference-in-Differences)**

This table replicates Table 3 using the Synthetic Difference-in-Differences (SDiD) estimator of Arkhangelsky et al. (2021) as a robustness check. The dependent variables are the logarithm of the number of establishments and employees in the Information and Communications and Construction industries, respectively. Standard errors are obtained from 1,000 bootstrap replications. ***, **, and * indicate 1%, 5%, and 10% significance, respectively.

| | (1) | (2) | (3) | (4) |
|---|---|---|---|---|
| | Information and Communications Industry | | Construction Industry | |
| | Y: Log (Establishment) | Y: Log (Employee) | Y: Log (Establishment) | Y: Log (Employee) |
| Policy*North | 0.1476** | 0.0802 | 0.3044* | 0.1333 |
| | (0.0731) | (0.0837) | (0.1606) | (0.1262) |
| Observations | 738 | 738 | 738 | 738 |

**Table A8: Policy Impact on Demographic Structure Changes in Northern Metropolis Area (Synthetic Difference-in-Differences)**

This table replicates Table 4 using the Synthetic Difference-in-Differences (SDiD) estimator of Arkhangelsky et al. (2021) as a robustness check. The dependent variables are the logarithms of the ratio of university-educated residents, the number of higher-income households, and the number of households with larger family sizes, respectively. We also control district-level median monthly household income as macroeconomic controls. Standard errors are obtained from 1,000 bootstrap replications. ***, **, and * indicate 1%, 5%, and 10% significance, respectively.

| | (1) | (2) | (3) |
|---|---|---|---|
| | Y: Log (Ratio_University) | Y: Log (Higher-income households) | Y: Log (Households with Larger Family Sizes) |
| Policy*North | 0.0721*** | 0.0871** | 0.0396** |
| | (0.0191) | (0.0348) | (0.0158) |
| HHIncome | YES | NO | YES |
| Observations | 126 | 126 | 126 |

**Table A9: Policy Impact on Increasing Housing Transaction Amount in Bordering Area in Shenzhen Housing Market**

This table presents the policy impact on aggregated transaction amount in the bordering area in Shenzhen housing market. $Log(Aggregated\ Transaction\ Amount)$ is defined as the logarithm of the building-month level transaction amount. $Policy$ indicates transactions post-October 2021 (i.e., after the announcement). $Border$ is also a dichotomous variable that equals 1 if the property is located within the bordering area in Shenzhen and 0 otherwise. In Columns (1) and (2), results include year-month and subdistrict fixed effects. In Columns (3) and (4), we also include year-month-by-district fixed effect. Standard errors are adjusted for clustering at year-month and district levels, are included in parentheses, and ***, **, and * indicate 1%, 5%, and 10% significance, respectively.

| | (1) | (2) | (3) | (4) |
|---|---|---|---|---|
| | Y: Log(Aggregated Transaction Amount) | | | |
| Variable | [-4, +4] Months | [-12 +12] Months | [-4, +4] Months | [-12 +12] Months |
| Policy*Border | 0.2065*** | 0.0737* | 0.1701** | 0.0289 |
| | (0.0718) | (0.0411) | (0.0762) | (0.0432) |
| Year-month FE | YES | YES | YES | YES |
| Subdistrict FE | YES | YES | YES | YES |
| Year-month*District FE | NO | NO | YES | YES |
| Observations | 2,837 | 9,094 | 2,837 | 9,094 |
| R-squared | 0.0720 | 0.1655 | 0.0805 | 0.1731 |

**Table A10: Policy Impact on Firm Establishments in Bordering Area in Shenzhen**

This table presents the policy impact on firm establishments in Shenzhen. $Log(Establishment)$ is defined as the logarithm of the number of establishments in quarter $t$ and subdistrict $l$. Column (1) and Column (2) show the results in information and communications industry and construction industry separately. $Policy$ is a dummy variable that equals 1 if the date of purchase is after October 2021 (i.e., after the announcement), and 0 otherwise. $Border$ is also a dichotomous variable that equals 1 if the subdistrict is located within the bordering area in Shenzhen and 0 otherwise. We have year-quarter and subdistrict fixed effects. Standard errors are adjusted for clustering at subdistrict levels, are included in parentheses, and ***, **, and * indicate 1%, 5%, and 10% significance, respectively.

| | (1) | (2) |
|---|---|---|
| | Information and Communications Industry | Construction Industry |
| | Y: Log (Establishment) | Y: Log (Establishment) |
| Policy*Border | -0.0656 | 0.0803 |
| | (0.1173) | (0.1379) |
| Year-quarter FE | YES | YES |
| Subdistrict FE | YES | YES |
| Observations | 2,856 | 3,375 |
| R-squared | 0.7543 | 0.2241 |